%% file: RF-Chord.tex
\documentclass[letterpaper,twocolumn,10pt]{article}

\usepackage{usenix, booktabs}
\usepackage{amssymb}
\usepackage{siunitx}

\newif\ifheadnice
\headnicefalse

\newcommand{\bo}[1]{{\color{blue}#1}}

\renewcommand{\bo}[1]{{#1}}

\newcommand{\revision}[1]{{\color{blue}#1}}
\renewcommand{\revision}[1]{#1}

\input{preamble}

\def\sys{\textsc{RF-Chord}\xspace}
\newcommand{\BULLET}{\vspace{+.02in} \noindent $\bullet$ \hspace{+.00in}}

\def\pku{\superscript{$\ast$}}

\begin{document}

\title{
\Large{\sys: Towards Deployable RFID Localization System for Logistics Network}
}

\def\pku{\superscript{$\mathbb{P}$}}
\def\ali{\superscript{$\mathbb{A}$}}
\def\ucsd{\superscript{$\mathbb{U}$}}
\def\mit{\superscript{$\mathbb{M}$}}
\def\hcs{\superscript{$\mathbb{K}$}}
\def\tsinghua{\superscript{$\mathbb{T}$}}

\author{
 Bo Liang\pku\ali, Purui Wang\mit, Renjie Zhao\ucsd, Heyu Guo\pku, Pengyu Zhang\ali, Junchen Guo\ali \\
 Shunmin Zhu\tsinghua\ali, Hongqiang Harry Liu\ali, Xinyu Zhang\ucsd, Chenren Xu\pku\hcs\superscript{\Letter}
\thanks{Bo Liang and Purui Wang are the co-primary student authors. Purui Wang was affiliated to Peking University and Alibaba Group during which he contributed to this work. \Letter: chenren@pku.edu.cn}
\\
 \vspace{0mm}
    \normalsize
    \begin{tabular}{cccc}
      {\pku}Peking University & {\ali}Alibaba Group &   {\mit}Massachusetts Institute of Technology & {\ucsd}University of California San Diego
    \end{tabular}\\
    \normalsize
    \begin{tabular}{cc}
      {\tsinghua}Tsinghua University & {\hcs}Key Laboratory of High Confidence Software Technologies, Ministry of Education (PKU)
    \end{tabular}\\
}


\maketitle
\input{00-abstract} 
\input{01-intro}

\input{011-table}
\input{02-system_overview}
\input{03-sys-oneshot}

\input{04-sys-longrange}
\input{06-sys-alg}

\input{07-impl}

\input{08-eval}

\input{09.1-discuss}

\input{03-related}

\input{09-concl}
\input{11-ack}
\newpage

\bibliographystyle{unsrt}
\bibliography{RF-Chord.bib}

\input{10-proofs}

\end{document}

%% file: preamble.tex
\usepackage[english]{babel}
\usepackage[utf8]{inputenc}

\newcommand{\COMMENTS}{yes}

\let\tempx\rmdefault
\usepackage{lmodern}
\let\rmdefault\tempx

\usepackage{color,balance,xspace,verbatim,ifthen,engord}

\usepackage{gensymb,amsmath,wasysym,amsthm,marvosym}
\usepackage{booktabs,colortbl,diagbox,multirow,tabularx,tablefootnote} 

\usepackage{dblfloatfix} 

\usepackage{subcaption} 
\usepackage{graphicx,epsfig,epstopdf,wrapfig}
\graphicspath{{./figures/}}
\DeclareGraphicsExtensions{.pdf,.mps,.png,.jpg,.eps,.PNG,.JPG}
\usepackage{algorithm,algorithmic}

\usepackage{url}
\newcommand{\secref}[1]{\S\ref{#1}}
\newcommand{\figref}[1]{Fig.~\ref{#1}}
\newcommand{\tabref}[1]{Tab.~\ref{#1}}
\newcommand{\eqnref}[1]{Eqn.~\ref{#1}}
\newcommand{\algref}[1]{Alg.~\ref{#1}}

\newcommand{\figcaption}[1]{\vspace{-8mm}\caption{#1}\vspace{-4mm}}

\usepackage{tikz}
\newcommand{\WoB}[1]{{\small \tikz[baseline=(char.base)]{\node[shape=circle,fill=black,draw,inner sep=0.5pt] (char) {\color{white}#1};}}} 

\newcommand*\circled[1]{\tikz[baseline=(char.base)]{\node[shape=circle,draw,inner sep=0.5pt] (char) {#1};}}

\newcommand{\superscript}[1]{\ensuremath{^{\textrm{#1}}}}
\newcommand{\argmax}{\operatornamewithlimits{argmax}}

\newcommand{\nosection}[1]{\vspace{3pt}\noindent\textbf{#1}}
\newcommand{\nosubsection}[1]{\vspace{3pt}\noindent$\bullet$\hspace{1mm}\textit{#1}}

\def\eg{\textit{e.g.,}\hspace{1mm}}
\def\ie{\textit{i.e.,}\hspace{1mm}}

\def\etc{\textit{etc.}\hspace{1mm}}

\usepackage{courier} 

\usepackage{enumitem}

\ifthenelse{\equal{\COMMENTS}{yes}}{
	\newcommand{\todo}[1]{\textcolor{red}{\textbf{TODO:} #1}}
	\newcommand{\fye}[1]{\textcolor{red}{#1}}  
	\newcommand{\remind}[1]{\footnote{\textit{\textcolor{red}{\textbf{Remind:} #1}}}}
	
	\newcommand{\del}[1]{\color{blue} {\sout{#1}}}
	\newcommand{\p}[1]{\vskip 1ex \noindent\colorbox{yellow}{\parbox{\columnwidth}{#1}}\vskip 4pt}
	\newcommand{\note}[1]{\vskip 4ex \noindent\colorbox{yellow}{\parbox{\columnwidth}{#1}}\vskip 6ex} 
	\newcommand{\q}[1]{\vskip 1ex \noindent\colorbox{magenta}{\parbox{\columnwidth}{\textbf{Question:} #1}}\vskip 4pt} 
	\newcommand{\qa}[1]{\hl{\textbf{Answer:} #1}}
}{
	\newcommand{\todo}[1]{}
	
	\newcommand{\fye}[1]{}
	\newcommand{\remind}[1]{}

	\newcommand{\del}[1]{}
	\newcommand{\p}[1]{}
	\newcommand{\note}[1]{}

	\newcommand{\q}[1]{}
	\newcommand{\qa}[1]{}	
}

\ifheadnice
\makeatletter
\renewcommand{\section}{\@startsection{section}{1}{\z@}{-.8ex \@plus -.4ex \@minus -.4ex}{.8ex \@plus .4ex \@minus .4ex}{\normalfont\large\bfseries\MakeTextUppercase}} 
\renewcommand{\subsection}{\@startsection{subsection}{2}{\z@}{-.8ex\@plus -.2ex \@minus -.2ex}{.4ex \@plus .2ex \@minus .2ex}{\normalfont\large\rmfamily\bfseries}} 
\renewcommand{\subsubsection}{\@startsection{subsubsection}{2}{\z@}{-.4ex\@plus -.2ex \@minus -.2ex}{.2ex \@plus .2ex \@minus .2ex}{\normalfont\large\slshape}}

\makeatother
\else
\fi



%% file: 00-abstract.tex
\begin{abstract}
RFID localization is considered the key enabler of automating the process of inventory tracking and management for the high-performance logistic network. A practical and deployable RFID localization system needs to meet {\it reliability}, {\it throughput}, and {\it range} requirements. This paper presents \sys, the first RFID localization system that simultaneously meets all three requirements. \sys features a multisine-constructed wideband design that can process RF signals with a 200 MHz bandwidth in real-time to facilitate one-shot localization at scale. In addition, multiple SINR enhancement techniques are designed for range extension. On top of that, a kernel-layer near-field localization framework and a multipath-suppression algorithm are proposed to reduce the 99th long-tail errors. Our empirical results show that \sys can localize up to 180 tags 6 m away from a reader within 1 second and with 99th long-tail error of 0.786 m, achieving a 0\% miss reading rate and \textasciitilde 0.01\% cross-reading rate in the warehouse and fresh food delivery store deployment.
\end{abstract}

%% file: 01-intro.tex
\section{Introduction}\label{sec:intro}
Today’s major e-commerce companies like Alibaba and Amazon need to handle a package volume that is tens of billions per year \cite{package_num}, calling for increasingly high-performance automated logistics operations in their network. Considering a typical warehouse in which tens or even hundreds of packages pass through each checkpoint -- the packages need to be verified, recorded, sorted, and tracked when checking in/out. 
\revision{In widely adopted barcode-based logistic networks, the worker spends 1\textasciitilde3 seconds on scanning one package. Although this operation can be automated by robots \cite{sorting}, the line-of-sight and field of view requirements of vision-based approaches limits work range and scalability fundamentally.}
RFID technology, since its invention, has been carrying the vision of replacing inefficient labor and automating inventory management with zero power, near-zero cost, and high throughput.

\begin{figure}
    \centering
    \includegraphics[width=\linewidth]{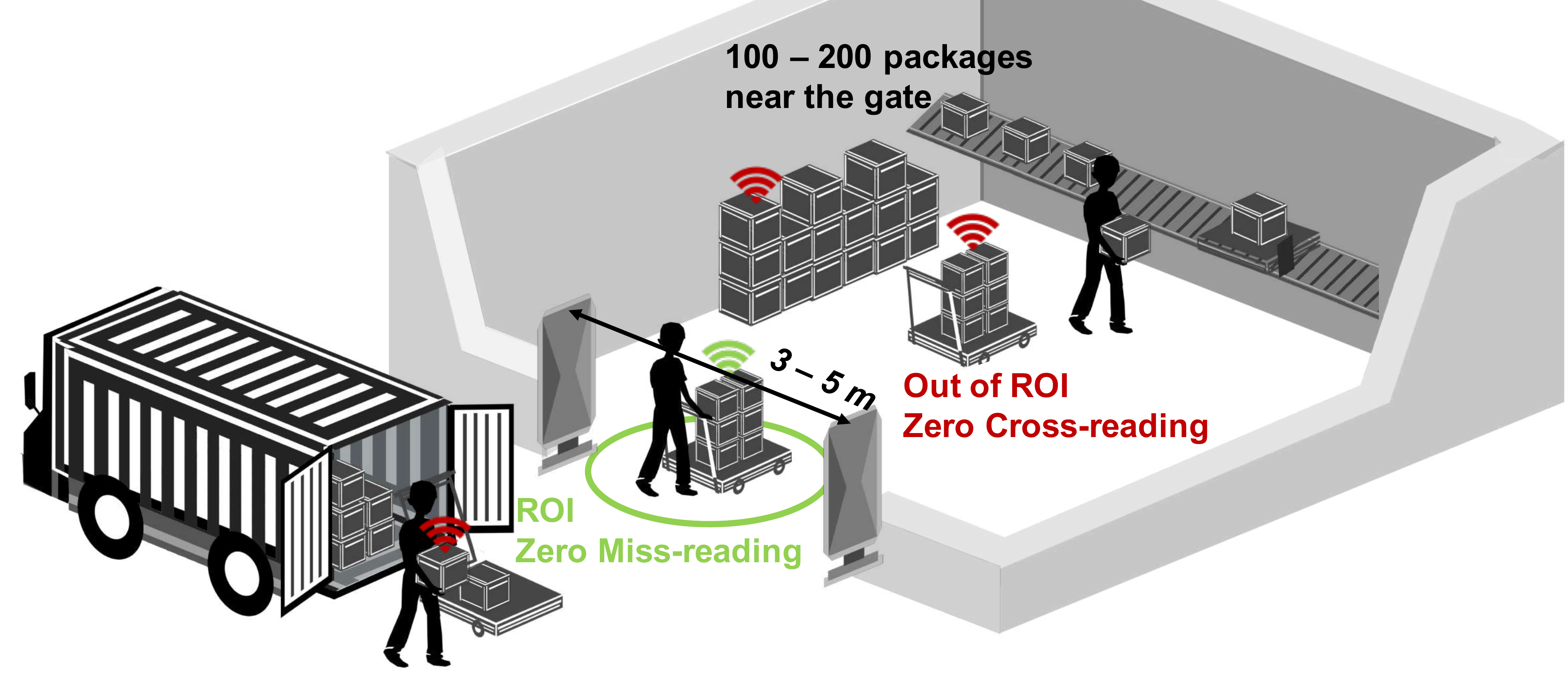}
    \vspace{-4mm}   
    \caption{In a typical logistic scenario, the packages are discharged from the truck, scanned at an inventory gate and sorted for warehouse check in. The RFID-based inventory gate should meet \textit{reliability}, \textit{throughput}, and \textit{range} requirements at the same time.}
    \vspace{-4mm}
    \label{fig:logistic}
\end{figure}

Towards a highly practical and deployable RFID empowered automated logistic network shown in \figref{fig:logistic}, there are three key considerations:
\textit{i) Reliability.} The classic ROI (range of interest) reading task requires the reader to scan all the RFID tags within the ROI (\ie near-zero miss-reading rate) while excluding any tag out of the ROI (\ie near-zero cross-reading rate);
\textit{ii) Throughput.} The packages come to the checkpoint in a burst (\ie 100\textasciitilde200 per pallet)
\footnote{Even though one trailer can carry up to 50 packages, the reader should be able to cover all the tags (100\textasciitilde200 tags) near the gate (including passed trailer and undischarged packages) to ensure to read all the passing packages.}
while all the logistic operations, including \bo{verification and recording} need to be finished within 2\textasciitilde3 seconds before check-in/out; 
\textit{iii) Range.} A single reader should cover tags within 3\textasciitilde5 m, which is the typical width of the check-in/out aisle.

Unfortunately, today's read-or-not inventory systems, both industrial products and research prototypes, all have limitations in meeting these three requirements simultaneously.
Industry-grade RFID systems (\eg Impinj) suffer from miss-reading and cross-reading when deployed in the logistic warehouses.
RFGo \cite{bocanegra2020rfgo} reports 99.8\% recall with 10 carrier-level synchronized antennas and neural network based classifier but limits its operating range to sub-meter. NFC+ \cite{zhao2020nfc+} achieves a sharp inventory boundary with magnetic resonance engineering that meets the reliability (\ie miss-reading rate of 0.03\% and cross-reading rate of 0\%) and range (\textasciitilde 3 m) requirements but cannot achieve the desired throughput. No current inventory-based solutions can support automatic package management in a practical logistics network. 

RFID localization technique offers an alternative approach toward the same goal by filtering out the reading outside the ROI.
Compared with the inventory-based system, the tag location brings a new dimension of information, which can realize a more flexible and accurate ROI reading.
The reliability of ROI reading depends on localization accuracy.
However, the legitimate narrow frequency band (\ie 26 MHz ISM band within 902\textasciitilde928 MHz) of RFID fundamentally limits its capacity of combating multipath and ambiguity \cite{li2011bandwidth}. 
To improve the localization accuracy, approaches like fingerprinting \cite{wang2013dude} and synthetic aperture radar (SAR) based hologram \cite{yang2014tagoram,shangguan2016design} have been proposed. However, they suffer from prolonged latency due to lots of tag inventory, especially at scale.
Cross-frequency based approaches utilize higher frequency band to overcome the bandwidth limitation (\eg 2.4 GHz \cite{ma20163d, an2018cross}, millimeter wave \cite{adeyeye2019miniaturized}, UWB \cite{arnitz2009multifrequency, decarli2015passive}) but introduce extra tag manufacturing cost due to wider frequency response and higher power attenuation.
More recently, sniffer-based RFID architecture \cite{ma2017minding,luo20193d} has been proposed to leverage the advantage of wideband (\eg 100\textasciitilde200 MHz) near 915 MHz to boost location accuracy without violating FCC regulation. Despite the potential, these systems either suffer from latency issues due to the lack of hardware support on multi-band parallel information capture \cite{ma2017minding}, or report limited sub-meter range \cite{luo20193d}.

\begin{table*}[]
\resizebox{\linewidth}{!}{
\begin{tabular}{lllll}
\toprule
\backslashbox{{\bf Solutions}}{{\bf Requirements}} & {\bf Throughput (> 100 tags/s)} & {\bf Range (> 3 m)} & {\bf Reliability (Near Zero Miss-reading $\&$ Cross-reading)} & {\bf Commercial Tag}\\ 
\midrule
Barcode (widely deployed) & No (\textasciitilde1 tag per second) & No (\textasciitilde1 m) & High (depend on the human labor) & Yes \\
xSpan \cite{impinj_xspan} (Inventory based)  & Yes (\textasciitilde185 tags/s with 142 mode) & Yes (\textasciitilde10 m) & Low (\textasciitilde 6\% miss reading and \textasciitilde 2\% cross reading) & Yes \\ 
RFgo \cite{bocanegra2020rfgo}    & No (TDMA-based)   & No (sub-meter)   & High (99.8\% recall)  & Yes         \\ 
NFC+ \cite{zhao2020nfc+}    & No data reported    & Yes (\textasciitilde3 m)   & High (0\% miss reading and \textasciitilde0.03\% cross reading)  & No         \\ 
PinIt \cite{wang2013dude}    & No data reported & Yes (> 5 m)        & Median (a few decimeters)  & Yes         \\ 
RF-IDraw \cite{wang2014rf}   & No data reported & Yes (> 5 m)       & Low (sub-meter)   & Yes\\ 
Tagoram \cite{yang2014tagoram}    & No (0.2 second for one tag)    & No (\textasciitilde2 m)     & Median (a few decimeters)  & Yes     \\ 
MobiTagbot \cite{shangguan2016design} & No data reported  & No (\textasciitilde1.5 m)       & High (a few centimeters)    & Yes   \\ 
NLTL tags\cite{ma20163d} & No (depend on switching) & No (\textasciitilde1 m)  & High (a few millimeters) & No \\ 
mmwave RFID \cite{adeyeye2019miniaturized} & No data reported & No data reported & Median (a few decimeters) & No \\ 
RFind \cite{ma2017minding}  & No (6.4 second for one tag)  & Yes (> 5 m) & High (a few centimeters) & Yes   \\ 
TurboTrack \cite{luo20193d} & No data reported & No (sub-meter)  &  High (a few centimeters)    & Yes\\ 
{\bf\sys (Our system)}   & {\bf Yes (180 tags/s)} & {\bf Yes (6 m)}  & {\bf High (0\% miss reading and \textasciitilde0.01\% cross reading)}   & {\bf Yes}\\ 
\bottomrule
\end{tabular}}
\vspace{-2mm}
\caption{Comparing \sys with state-of-the-art wireless systems for logistic network requirements.}
\vspace{-2mm}
\label{tab:compare}
\end{table*}

\begin{figure*}[h]
    \centering
    \includegraphics[width=\linewidth]{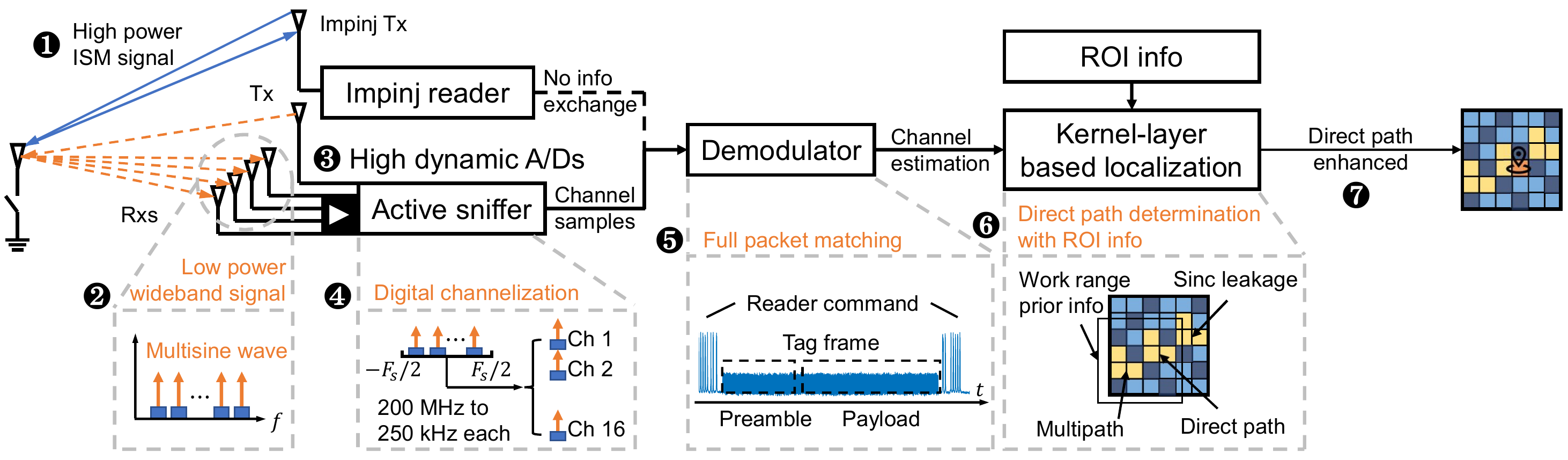}
    \vspace{-6mm}   
    \caption{\sys system overview.}
    \vspace{-3mm}
    \label{fig:sys_overview}
\end{figure*}

This paper introduces the design and implementation of \sys, an active sniffer-based wideband RFID localization system that tackles the above challenges. 
\sys exploits wideband signal and a hologram-based localization algorithm to realize {\it high reliability}. It employs lossless data stream compression and a GPU-based decoder to guarantee real-time decoding and channel estimation for {\it high throughput}. It utilizes a customized wideband waveform, full packet matching integration, fine-grained clock offset mitigation, and channel diversity decoding to improve SINR for {\it long range}.

\revision{
\sys ensures {\it high reliability} (\ie near zero miss reading and cross reading) by high-accuracy localization.
Our study (\secref{ssec: error source}) shows that the multipath profile causes long-tail localization errors. Therefore, we design the fine-grained distance resolution hardware and multipath-suppression algorithm to handle these long-tail errors. 
Considering that the distance resolution is inversely proportional to bandwidth (\ie $\frac{c}{2B}$), the distance resolution of a conventional UHF RFID reader, which works on a 26 MHz wide ISM band, is only 5.78 m. \sys introduces an extra active sniffer-based reader to help UHF RFID reader realize 200 MHz parallel wideband localization (\secref{ssec: wideband backscatter}).
However, the distance resolution of 200 MHz (0.75 m) is still not enough in all situations. \sys exploits a kernel-layer-based near-field localization algorithm framework to handle corner cases. 
The kernel function characterizes the location estimation from a single channel, and layer functions coherently combine multiple channels into a final location estimation.
This framework supports choosing different kernel and layer functions suitable for various deployment scenarios to achieve multipath suppression and ambiguity reduction (\secref{hologram}). For example, in \sys's deployment in the warehouse, the work range is fixed so it can be taken as prior information for direct path enhancement to effectively suppress the multipath effect (\secref{ssec: sys kernel and layer}).

\sys ensures {\it high throughput} by one-shot channel measurement and one-shot location estimation.
The hardware supports concurrent phase and amplitude capture across multiple antennas and wide bandwidth. Therefore, \sys can obtain the necessary information (\ie wideband channel estimation across multiple antennas) for localization within only one shot measurement. 
It is challenging because:
i) directly capturing the wideband signal from a large array will result in a huge amount of real-time data (\textasciitilde64 Gbps);
ii) the commercial reader does not support real time synchronization  (\ie synchronizing with our sniffer-based reader at each slot \cite{uhfc1g2}).
Utilizing the essence that the wideband backscattered signal is a combination of scattered narrowband signals, \sys distills 4 MHz valid bandwidth from 200 MHz bandwidth to reduce the data rate by 50x without information loss (\secref{ssec:channelization}). 
Meanwhile, we develop a GPU-based wideband decoder to ensure real time decoding and channel estimation. In other words, the sniffer-based reader has an independent decoder and does not depend on any specific commercial reader interface.
It makes our design adaptive to any ISM band commercial reader, which primarily serves as a power activator and multiple access handler (\secref{sec:improve_SINR}).
Finally, \sys supports one-shot localization with 8 antennas and 16 frequencies across 200 MHz in \textasciitilde5 ms.


\sys ensures {\it long range} (up to 6 m) with multi-sine waveform sniffer and sophisticated wideband channel information estimation. To follow the FCC regulation, the strength of the sniffer excitation signal needs to be {\it smaller than -13.3 dBm} (see \secref{sec:FCC} for the calculation), which is 50 dB weaker than that of commercial readers. \sys features the following designs for signal-to-interference-plus-noise ratio (SINR) enhancement without modifying the tag chip: 
i) It exploits a multisine waveform, which constructs a whole 200 MHz band by taking samples with multiple narrow bands, to significantly reduce the noise bandwidth (\secref{ssec:reduce_BW});
ii) It handles the high dynamic range requirements introduced by self-interference through high-resolution digital channelization and a low crest factor waveform design (\secref{ssec:self_inter}); 
iii) It further exploits the integration gain of full packet matching (\secref{ssec: full packet matching}) and performs accurate tag clock offset mitigation (\secref{ssec:chan_est}) and decoding with channel diversity (\secref{ssec:joint decoder}).}

We deploy \sys and our results show that \sys presents the first RFID (localization) system meeting all the requirements (\ie reliability, throughput, and range) in the logistic network (\tabref{tab:compare}). The key results are:

\BULLET \textbf{Reliability.} We evaluate \sys's performance at 384 locations and collect over 20k tag responses in the lab environments. Its 99\% localization error is 0.786 m. 
We deploy \sys in the dock door of a warehouse and the scanning gate of a fresh food delivery store. We find that it could read 100\% of the tags passing the checkpoint (0\% miss-reading rate). Its cross-reading rate is only 0.0025\%\textasciitilde0.0154\%, which is up to 12x improvement compared to state-of-the-art \cite{bocanegra2020rfgo,zhao2020nfc+}. 

\BULLET \textbf{Throughput.} \sys can localize up to 180 tags per second, which is very close to pure inventory devices \cite{impinj_xspan} and two to three orders of magnitude faster than state-of-the-art localization systems \cite{ma2017minding,yang2014tagoram}.

\BULLET \textbf{Range.} \sys can localize tags 6 m away from the reader with transmit power below -15 dBm. There is no obvious throughput and reliability loss with distance increasing.


We open sourced the \sys's hardware and software as well as the evaluation dataset in \url{https://soar.group/projects/rfid/rfchord}.

%% file: 011-table.tex
%% file: 02-system_overview.tex
\section{\sys's System Overview}\label{sec:system_overview}
A high level operational flow of \sys is shown in \figref{fig:sys_overview}. \sys embraces any ISM-band reader \WoB{1} as the tag activator that is capable of charging, coordinating multiple access over EPC Gen II tags. Active sniffer reader observes tags by emitting a low power (-15 dBm) wideband multi-sine waveform to pick up tag responses over a wide frequency band.
Specifically, we build the RF frontend and FPGA hardware \WoB{2} as a scalable platform that can receive the tag response from 8 antennas and 16 frequencies of carriers simultaneously. Furthermore, despite the strict legal emission power limit, we still achieve a long range in sniffing the tag response in the wideband without exchanging any information (\eg EPC ID) with the ISM-band reader. \sys achieve independent decoding and channel estimation by using dynamic range optimization \WoB{3}, digital channelization \WoB{4} in hardware, and a real-time full packet matching \WoB{5} in software. After one-shot tag inventory, \sys obtains adequate information from both frequency and spatial domains, which are important for robust localization in a multipath-rich environment. 
\sys also uses a kernel-layer-based near-field localization algorithm to suppress the multipath effect. This algorithm identifies the direct path with the time of flight profile and prior knowledge (region of interest or ROI information in our paper) \WoB{6}. Then it enhances the direct path and estimates the location with a summation layer (a form of near-field AoA+ToF localization) \WoB{7}.

%% file: 03-sys-oneshot.tex
\section{One-shot Wideband with Multisine Wave}\label{ssec:multisine}
This section explains why we select multisine wave as the wideband signal and how \sys acquires fine-grained tag responses in one shot. We review the primer of the backscatter signal model and its fundamental narrowband constraint. Then we present our design of constructing a wideband backscatter signal with the multisine waveform on Tx and slicing it for real-time parallel processing on Rx.

\subsection{Backscatter Signal Model Primer}

The basic backscatter operation in RFID systems is shown in \figref{fig:bs_model}. A device emits a high-power single-tone excitation signal $s(t)$ to power the tag and act as a carrier. This carrier will be modulated by the baseband signal $B_{\mathrm{tag}}(t)$ of the tag. The resulting (mixed) backscattered signal is:
$$r(t) = s(t) \cdot B_{\mathrm{tag}}(t)$$ 
Note that the bandwidth of $r(t)$ is the summation of that of s(t) and $B_{\mathrm{tag}}(t)$, and $B_{\mathrm{tag}}(t)$ is typically a narrowband signal\footnote{We take 250 kHz as the bandwidth $B_{\mathrm{tag}}(t)$ for the whole paper according to the standard \cite{uhfc1g2}.} for low power purpose according to the EPC Gen II standard. Therefore, the backscattered signal $r(t)$ will also be narrowband given that $s(t)$ is a single tone.

\subsection{Backscattering with Wideband Multisine Wave}\label{ssec: wideband backscatter}
\begin{figure}
	\centering
	\begin{subfigure}{0.49\linewidth}
		\includegraphics[width=0.9\linewidth]{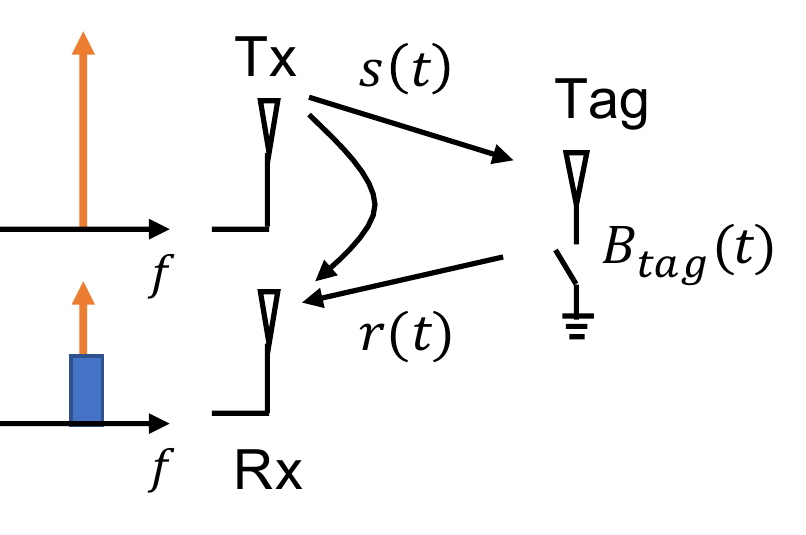}
		\caption{\label{fig:bs_model} Single-tone Backscatter.}
		\vspace{1mm}
	\end{subfigure}
	\begin{subfigure}{0.49\linewidth}
		\includegraphics[width=0.9\linewidth]{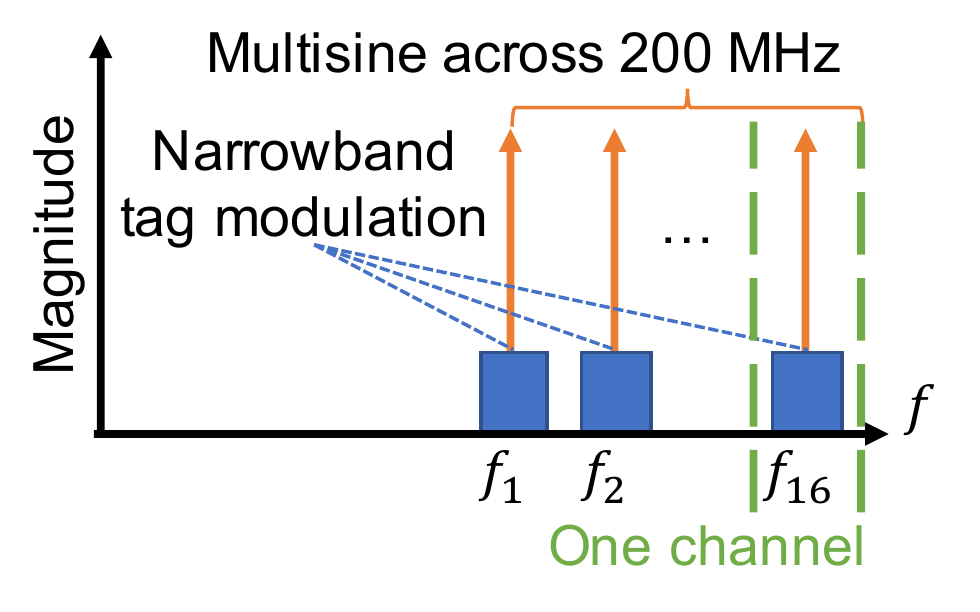}
		\caption{\label{fig:multisine_spect} Multisine Excitation Signals.}
		\vspace{1mm}
	\end{subfigure}
	\vspace{-3mm}
	\caption{Model of Multisine Backscatter.}
	\vspace{-5mm}
\end{figure}

When applying a wideband signal $s(t)$, one can retrieve a wideband backscatter signal $r(t)$.
Following this idea, \sys adopts a multisine signal as $s(t)$. The multisine signal is a combination of multiple single tones across wide band with the same amplitude $s(t)=\sum_i \sin(f_it+\phi_i)$. The backscattered signal will be $r(t) = \sum_i B_{\mathrm{tag}}(t) \cdot \sin(f_it+\phi_i)$. \sys adopts 16 carriers with different frequencies across a 200 MHz band in the practical implementation. 
\figref{fig:multisine_spect} shows the spectrum of multisine signal $s(t)$ with backscatter signal $r(t)$. Since the difference between each carrier frequency is much larger than the bandwidth of $B_{\mathrm{tag}}(t)$, the received signal can be treated as multiple copies of $B_{\mathrm{tag}}(t)$ modulated on different carrier $f_i$. 
Therefore, on Rx, $r(t)$ can be sliced to 16 individual narrowband channels without information loss, and then the channel information at each carrier frequency $f_i$ can be extracted by using a well-explored RFID processing mechanism (\eg mixing and demodulating) in parallel. 
In a nutshell, we sample the wideband with multiple narrowband signals, enabling \sys to construct the wideband channel information within one shot.

\subsection{Why Multisine Wave}\label{ssec:why multisine}
The multisine waveform has two advantages. First, the multisine waveform is adaptive to conventional narrowband decoding and channel estimation because the signal in each channel is still narrowband. Extracting these narrowband signals can achieve excellent data rate compression (\secref{ssec:channelization}). Second, the multisine waveform is amenable to noise and interference reduction because of the low noise bandwidth and low chances of being interfered with, resulting in SINR enhancement, which improves the work range (\secref{ssec:reduce_BW}).
Compared with the two alternative well-known wideband waveform choices, frequency hopping \cite{ma2017minding} and OFDM signal \cite{luo20193d}, the multisine waveform is more efficient because it avoids the time overhead in switching between carriers introduced by the former one, and uses the same bandwidth as the (tag) modulation bandwidth, which is 250 kHz out of the full 200MHz bandwidth used by the latter one. 
In fact, this wideband but narrow sample signal can introduce 29 dB gain on the SINR compared to the full wideband signal (see \secref{ssec:reduce_BW}), which means around 5$\times$ range under the same transmit power. Furthermore, since the multisine wave captures all the backscatter signals in the time domain, the whole packet of tags can be fully utilized for integration gain to improve the SINR (see \secref{ssec:chan_est}).

\begin{figure}
	\centering
	\begin{subfigure}{0.53\linewidth}
		\includegraphics[width=\linewidth]{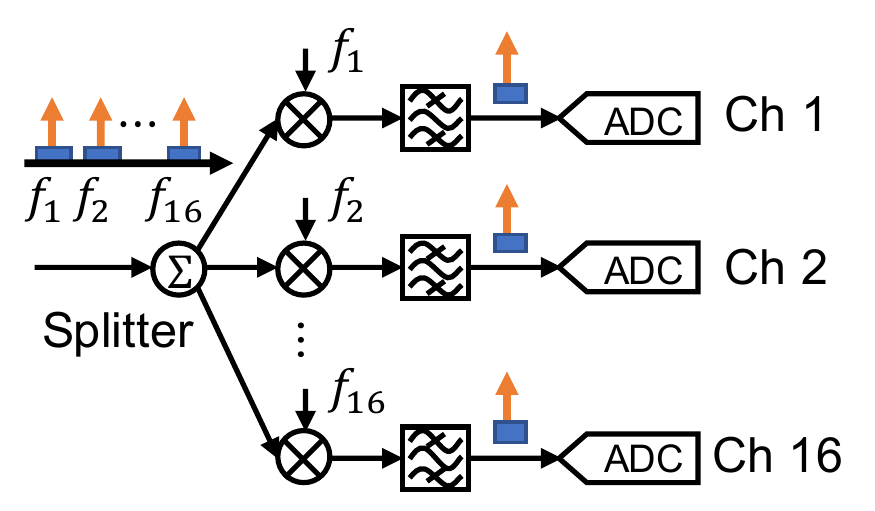}
		\vspace{-5mm}
		\caption{Analog Channelization.}
		\vspace{1mm}
		\label{fig:chan_analog}
	\end{subfigure}
	\begin{subfigure}{0.46\linewidth}
		\includegraphics[width=\linewidth]{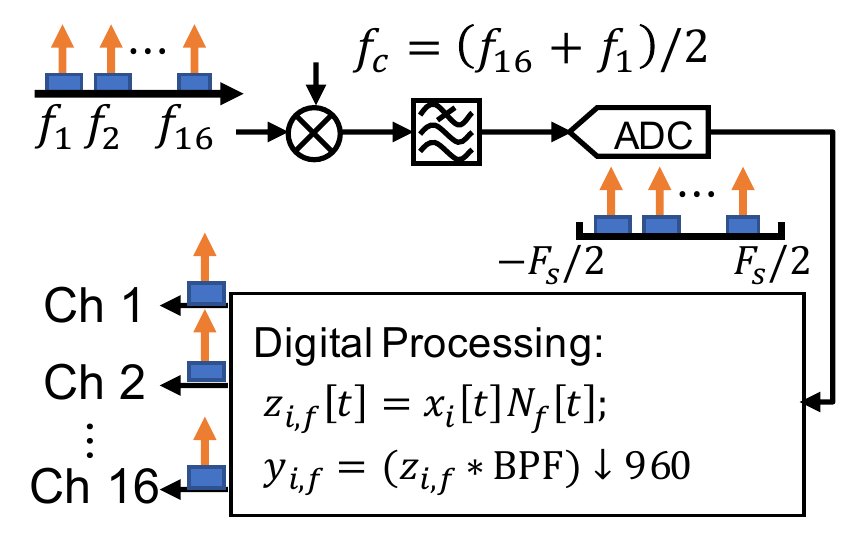}
		\vspace{-5mm}
		\caption{Digital Channelization.}
		\vspace{1mm}
		\label{fig:chan_digital}
	\end{subfigure}
	\vspace{-2mm}
	\caption{Two channelization approaches.}
	\vspace{-5mm}
	\label{fig:channelization}
\end{figure}

\subsection{Digital Wideband Channelization}\label{ssec:channelization}

\sys utilizes channelization, which enables one-shot capturing of wideband signals across multiple antennas and reduces the amount of data to be processed during real-time operation.
Channelization is a process of extracting effective narrowband signals from a received signal.
When a wideband tag signal is received, the aggregated bandwidth of 8 antennas will be 1.6 GHz, resulting in a total of 64 Gbps data (16-bit IQ sample, $1.25\times$ Nyquist). It is challenging to process such massive data in real time. However, recall that with a multisine excitation signal, the effective tag signal is only located around the carrier frequencies, as shown in \figref{fig:multisine_spect}. Therefore, the effective bandwidth of the system should be 8 $\times$ 16 $\times$ 250 kHz = 32 MHz, only 1/50 of the full 200 MHz bandwidth, so that channelization can compress the data validly without information loss. 

There are two channelization schemes to extract these narrowband signals: analog channelization and digital channelization. 
As shown in \figref{fig:chan_analog}, the sniffer with analog channelization has multiple RF chains for the corresponding channels. Each RF chain uses one carrier frequency $f_i$ as its local oscillator (LO) for down-conversion and a filter at the baseband to filter the signal from other channels out. Alternatively, digital channelization finishes all the aforementioned functions in the digital domain as shown in \figref{fig:chan_digital}. 

\sys adopts digital channelization -- the sniffer will generate and capture the whole multisine wave with one RF chain. On the Rx side, an ADC/DAC with a 245.76 MHz sampling rate captures all tag signals simultaneously. Further channel extraction can be achieved by digital down-conversion and digital filtering.
Digital channelization offers two significant benefits over analog channelization: 
First, it has better scalability because it only needs one RF chain for each antenna, regardless of the number of channels (and sine tones) are required, while in analog channelization, each channel needs an exclusive RF chain with bulk components (\eg mixer, PLL, and VCO).
Second, it is precisely synchronized among different tones in the multisine wave, while analog channelization needs extensive engineering efforts to synchronize among a large amount of ADCs/DACs and LOs.
Nevertheless, analog channelization still has it own advantages, including the convenience of extending or switching bandwidth by changing the carrier frequency and the lower requirements of ADC bandwidth.
\sys also embraces these advantages through the high-speed ADC and low crest factor multisine waveform, which will be introduced in \secref{sec:improve_SINR}.

%% file: 04-sys-longrange.tex
\section{SINR Improvement for Long Range} \label{sec:improve_SINR}
This section first presents how \sys improves SINR under long work range by reducing the external noise and canceling self-interference. It next explains how \sys exploits the full tag packet to incorporate the integration gain, which is based on the multiple channel decoder with clock offset mitigation.

\subsection{External Noise Suppression}\label{ssec:reduce_BW}
To follow the FCC regulation, the signal strength of each frequency component in the multisine is -15 dBm, which is 51 dB lower than the 36 dBm excitation signal in the ISM band (see details in \secref{sec:FCC}). With the low signal strength limitation but the long range requirement, we need to reduce the external noise and interference as much as possible.

\sys adopts the tag signal with reduced bandwidth for lower chances of in-band interference and lower noise. The relationship between thermal noise $P_{\mathrm{noise}}$ and signal bandwidth $B$ at room temperature can be expressed as $P_{\mathrm{noise}} = -174 + 10\mathrm{log}_{10}(B)$ \cite{TNeq}.
As described in \secref{ssec:channelization}, the digital channelization at the receiver separates a combined 200 MHz wideband signal into multiple 250 kHz narrowband signals. 
This means that the thermal noise can be reduced from -91 dBm to -120 dBm (29 dB gain).
Furthermore, the reduced bandwidth also reduces the probability of being interfered with by other devices working in the same band.

\begin{figure}
	\centering
	\begin{subfigure}{0.49\linewidth}
		\includegraphics[width=\linewidth]{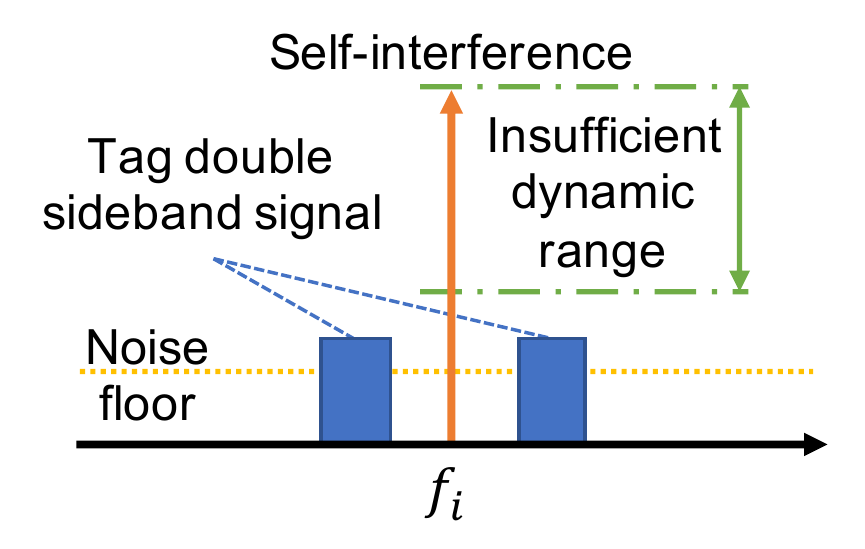}
		\vspace{-5mm}
		\caption{Dynamic Range.}
		\vspace{1mm}
		\label{fig:selfinter_DR}
	\end{subfigure}
	\begin{subfigure}{0.49\linewidth}
		\includegraphics[width=\linewidth]{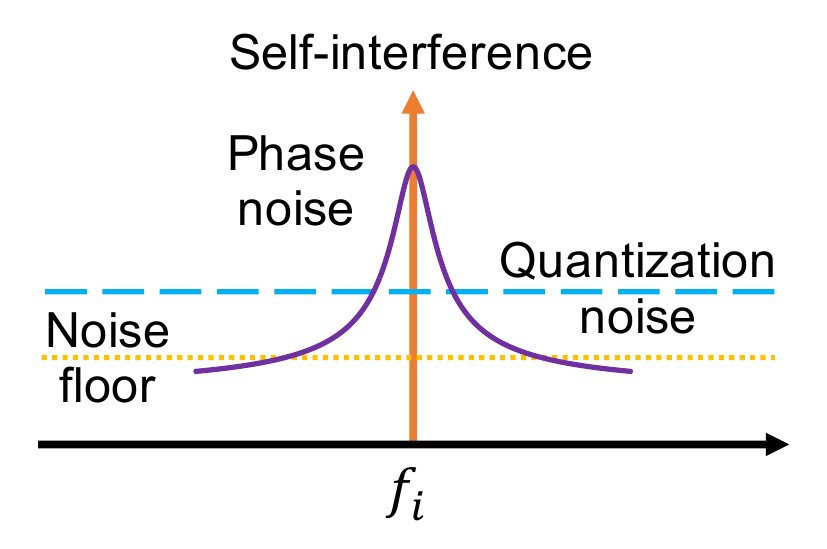}
		\vspace{-5mm}
		\caption{Unpredictable Noise.}
		\vspace{1mm}
		\label{fig:selfinter_noise}
	\end{subfigure}
	\vspace{-3mm}
	\caption{Two issues caused by self-interference.}
	\vspace{-4mm}
	\label{fig:self_inter}
\end{figure}

\subsection{Self-interference Canceling}\label{ssec:self_inter}
Besides the external interference from other devices, the self-interference caused by the natural full-duplex operation of our active sniffer will also limit the SINR. 
\sys's multisine waveform and low power transmission reduce the complexity of self-interference cancellation.
As shown in \figref{fig:selfinter_DR}, the self-interference in one channel is just a single tone after channelization. A commercial tag uses double-sideband modulation with a subcarrier to differentiate the tag signal from the single-tone excitation signal. Therefore, \sys uses filters to cancel the self-interference caused by the single tone.

However, given the wideband signals are too weak (\ie -15 dBm ), there remain two practical challenges. First, the dynamic range of the receiver may not be large enough to detect the tag signal. Second, any unpredictable noise, such as phase noise and circuit noise from Tx, will be transmitted along with the $s(t)$ and may bury the wideband signal. Then we'll go over how to deal with these issues.

\nosection{Dynamic Range.}
Dynamic range is the ratio between the largest and smallest values that the received signal can assume. Specifically, the largest value is the self-interference, and the smallest signal is the targeted wideband tag signal.  
As shown in the \figref{fig:selfinter_DR}, even though the tag signal strength is higher than the noise floor and interference, it can still be buried if the dynamic range is not large enough. 
\sys meets the requirement of dynamic range by adopting the following strategies: First, it adopts a high-resolution ADC because the dynamic range of the receiver will be bottlenecked by the dynamic range of the ADC. The theoretical dynamic range of the receiver is 6.02 N + 1.76 dB \cite{DReq}, where N is the resolution of the ADC. Therefore, a fundamental way to solve the issue is to increase the resolution of the ADC. \sys adopts 16-bit ADC, which has the largest resolution in 2022 when satisfying the 200 MHz bandwidth requirement.
Secondly, it adopts a carefully designed low crest factor multisine wave on the transmission side to relax the dynamic range requirement of the Rx. 
The intuition behind this is that since the dynamic range requirement on the ADC is more related to the peak amplitude of the self-interference signal instead of the average signal power, it can be relaxed by using a lower peak signal while remaining the average power.
The crest factor is the peak amplitude divided by the RMS value of the waveform, and for a multisine signal, it has been well studied that the crest factor can be reduced by tuning the phases $\phi_i$ in the multisine signal. 
Following the methods mentioned in \cite{Yang2015multisine}, the crest factor of the multisine waveform adopted by \sys can be reduced from 4 to 1.24 (or peak-to-average power ratio from 12 dB to 1.87 dB).

\nosection{Unpredictable Noise.}
The unpredictable noise is caused by the response of self-interference in the circuit.
As illustrated in \figref{fig:selfinter_noise}, the noise floor may be dominated by the phase noise, DAC quantization noise, \etc along with the self-interference. 
Fortunately, \sys does not require a dedicated cancellation circuit like \cite{van2017ADI_reader} because the power of \sys's self-interference is much lower than that of a commercial RFID reader. 
Moreover, \sys utilizes Analog Devices ADRV9009 transceiver of 16-bit ADC \cite{adrv9009} and HMC7044 VCXO-based clock tree \cite{HMC7044}, ensuring an optimal quantization and clock phase noise below the noise floor.
Therefore, the RF frontend of \sys's receiver is not saturated, and the noise will only go through the air instead of the feedback path of the receiver. The noise experienced by \sys is not dominated by unpredictable noise.

\subsection{Full Packet Matching}\label{ssec: full packet matching}
\sys estimates each channel in parallel and then combines them into a wideband channel estimation.
The standard channel estimation techniques for one channel can be expressed as follows:
$$h_i = \sum_t r(t)\hat{I}^*(t)$$
where $r(t)$ is received tag response and $\hat{I}(t)$ is a template. In most RFID systems, only the pilot signal part (RN16) is used for clock and phase estimation, and the main part of the tag signal (EPC ID) is left unused. \sys utilizes the full packet signal, including RN16 and EPC ID. The length of the signal will be extended from 0.31 ms to 2.31 ms when assuming the backscatter link frequency (BLE) of the tag is 250 kHz and the EPC ID length is 96 bits \cite{MonzaR6}. By doing the full packet matching, \sys can achieve  {$10\mathrm{log}_{10}{\frac{2.31}{0.31}}=8.7$ dB} integration gain. 

We need to generate a noiseless template of the full packet for full packet channel estimation. However, unlike the predefined pilot signal, the template of the packet changes depending on the tag's EPC ID. Collecting EPC ID and timestamp from a commercial reader device in real-time is unsupported due to the interface limitation: 
i) the available interface from a commercial reader is usually done by using asynchronous communication, which hinders real-time processing; 
ii) the timing information is usually not reported by commercial readers.
Therefore, \sys needs to decode the wideband signal into EPC ID independently.

\subsection{Clock Offset Mitigation}\label{ssec:chan_est}
\begin{figure}[t]
    \begin{centering}
        \includegraphics[width=\linewidth]{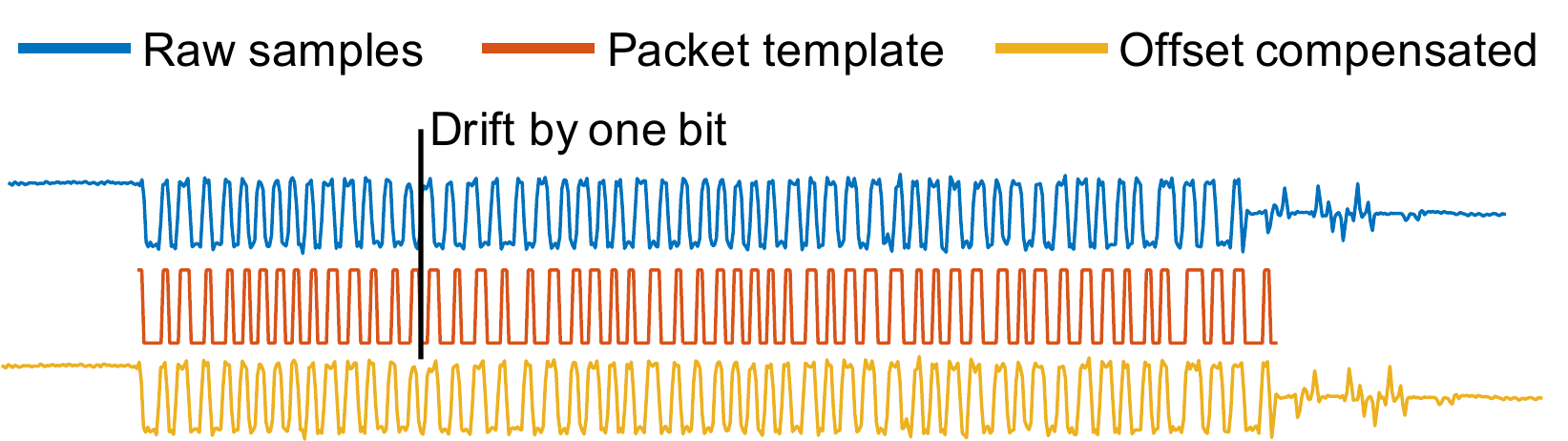}
    \vspace{-3mm}
    \par\end{centering}
    \caption{\label{fig:clock_offset} The waveforms of the tag signal with clock offset, the reference, and recovery signal from the offset.}
   \vspace{-4mm}
\end{figure}

Accurate decoding needs to mitigate the clock offset of the RFID tag signal.
Specifically, the protocol tolerates up to $\pm$ 10\% frequency offset and $\pm$ 2.5\% frequency fluctuation during backscattering (refer to Tab. 6.9 of \cite{uhfc1g2}).
For example, say we read a tag that is 2.5\% faster than nominal BLF. For a typical randomized uplink packet of 128 bits with a perfect match at the start of the frame, the received signal will be ahead of the template by one bit at the 32nd bit, and the remaining 96 bits thereafter contribute useless fluctuations to channel estimation, as figured out in \figref{fig:clock_offset}.
\sys needs to analyze the clock and estimate the offset parameters for mitigation, which can be described by:
$$\tau(t)= \mathrm{Square}((f_{\mathrm{BLF}}-\alpha_0-\alpha(t))(t-t_0))$$
Where $t_0$ is the actual start of frame (SOF), $\alpha_0$ is the initial clock frequency offset (CFO) from prescribed BLF, and $\alpha(t)$ is the fluctuation of the clock. Next, we introduce \sys's components which estimate these parameters.

\begin{figure*}[t]
\begin{minipage}[b]{0.35\linewidth}
    \begin{centering}
        \includegraphics[width=\linewidth]{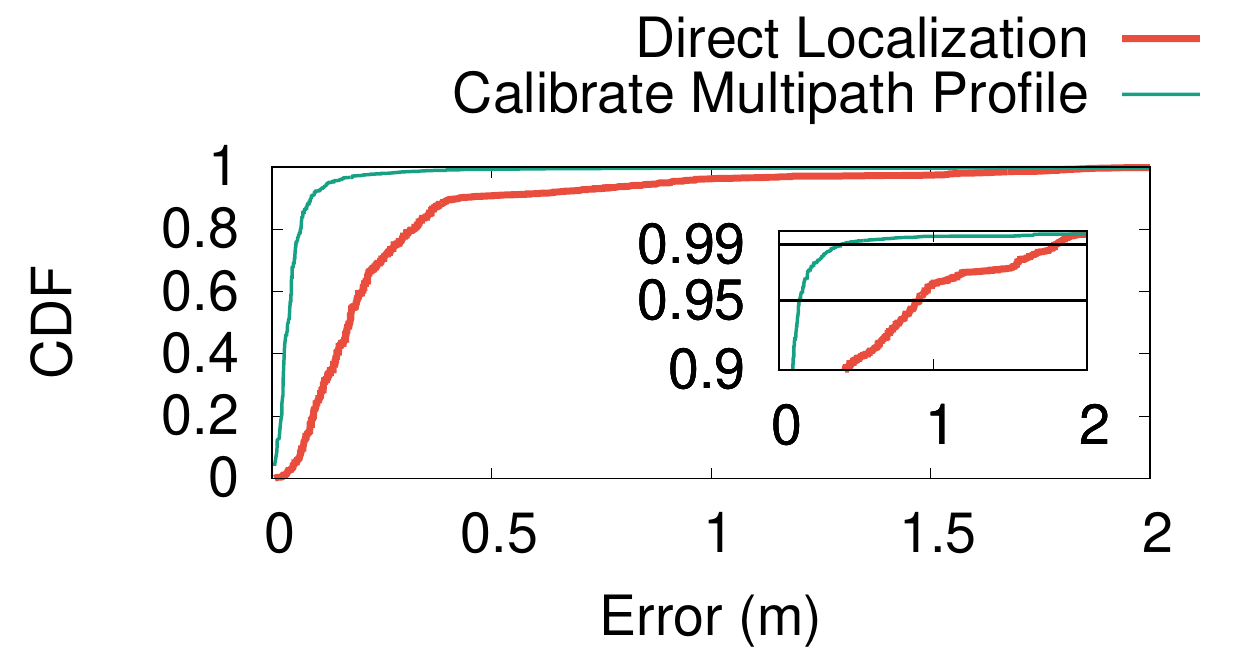} 
    \par\end{centering} 
    \vspace{0.5cm}
    \figcaption{\label{fig: error source} Eliminating the multipath effect reduces the 99th long-tail error.}
\end{minipage}
\hspace{2mm}
\begin{minipage}[b]{0.63\linewidth}
\centering	
    \begin{centering}
        \includegraphics[width=\linewidth]{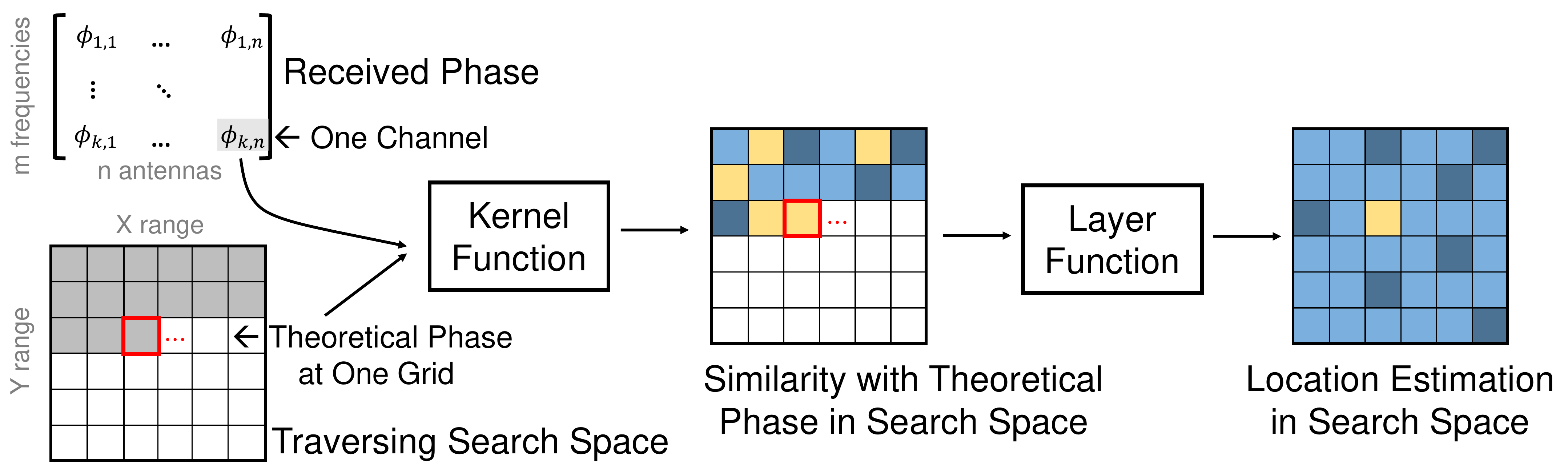} 
    \par\end{centering}
    \vspace{8mm}
    \figcaption{\label{fig: kernel layer framework} Kernel-Layer Framework.}
\end{minipage}
\end{figure*}

\nosection{Preamble Matching for $t_0$ and $\alpha_0$.} 
\sys first estimates the $t_0$ and $\alpha_0$ by adopting a standard sliding window correlator with a known preamble $p(t)$. Specifically, we derive the initial estimation of $\hat{t_0}$ and $\hat{\alpha_0}$ by this correlation calculation, where the $x(t)$ is the received samples, $p_\alpha(t)$ is the reference template tuned to a clock frequency of  $f_{\mathrm{BLF}} - \alpha_0$:

$$\{\hat{t_0}, \hat{\alpha_0}\} = \argmax_{t_0,\alpha_0}\left\vert\int_0^{T_p}p_{\alpha_0}^*(t)x(t+t_0)\mathrm{d}t\right\vert$$

\nosection{PLL to Track $\alpha(t)$ Variation.}
After eliminating $\alpha_0$, the clock still has residual offset $\alpha(t)$, which comes from the tag clock fluctuation during the communication and may be significant in the long packet.
Because the Miller code of RFID \cite{uhfc1g2} is a self-clocked and modulated bandpass signal, \sys can extract the subcarrier of the line code to track the clock frequency offset accurately. \sys adopts a feedback-based digital Costas PLL \cite{proakis} to track the clock continuously.

After compensating estimated clock $\tau(t)$, the clock offset is mitigated (the last waveform shown in \figref{fig:clock_offset}). We can see that the signal is well synchronized with the template.

\subsection{Decoding with Channel Diversity}\label{ssec:joint decoder}
After clock offset mitigation, we can decode the full packet, extract the correct template $\hat{I}(t)$, and assemble the decoder. 
Because the tag baseband signals on all channels are the same, \sys can apply nulling and beamforming algorithms to utilize the diversity across frequencies and antennas to make a joint decoder.
\sys combines the signals from all channels into one \textit{steered} single-channel signal --
it first performs an adaptive maximum signal-to-noise ratio (MSNR) beamforming over the array of each frequency to null the major jammer in the spatial domain and then performs maximum-ratio combining (MRC) beamforming across the frequency domain to improve the SINR further.
With this cleaned steered single channel, \sys exploits a Viterbi decoder to decode the EPC ID. It then applies the EPC ID to make accurate channel estimations on all the channels. 
A series of efforts introduced in this section, including suppressing external noise, canceling self-interference, matching full packet, mitigating clock offset, and decoding with diverse channels, guarantees \sys to extract wideband channel estimations at a long distance even with the ultra-low power emission signal.

%% file: 06-sys-alg.tex
\section{Localization with Kernel-Layer Framework}\label{sec: localization algorithm}

In this section, we first conduct empirical experiments which show: i) multipath is the primary factor that confines the
long-tail performance of the RFID localization system once the tag is successfully inventoried; ii) 200 MHz bandwidth
is not sufficient to eliminate all the long-tail errors caused by multipath. To address these problems, we propose a kernel-layer framework for localizing RFID tags in the near field. It can suppress long tail errors from multipath by enhancing the direct path and incorporating prior knowledge from logistics.

\subsection{Long-tail Errors Source Demystification}\label{ssec: error source}
We conduct a validation experiment to confirm that multipath is the primary source of long-tail localization errors. In this experiment, we put five tags at a distance of 4 m from the reader. We use 16 carriers evenly spaced across 200 MHz bandwidth, 8 antennas, and a hologram-based localization algorithm (see details in \secref{ssec:hologram primer}). There is a metallic heater 1.5 m from the tag as the multipath source. \figref{fig: error source} shows that the 99th localization error (red line) is 1.798 m, too large to ensure reliable usage in industry settings.
The theoretical analysis explains this observation -- the 200 MHz bandwidth is only able to differentiate paths that have a propagation distance difference larger than $c/(2B) \approx (3\times 10^8\text{ m/s})/(2\times 200\text{ MHz}) = 0.75$ m. Once the propagation distance of two paths is smaller than 0.75 m, which is common for many indoor deployments, 200 MHz is insufficient for differentiating one from the other.

Then we evaluate the performance without the multipath effect to check our results double. We keep the experiment setup, conduct RF measurement of a reference tag close to target tags and extract its phase offset from the groundtruth. Considering that the multipath profiles of nearby tags are similar, we subtract each tag's channel estimation with the offset from the reference tag.
The 99th localization error of the same set of tags decreases to 0.400 m (green line in \figref{fig: error source}). It proves that multipath is the primary factor determining the long-tail performance of the RFID localization system, even with 200 MHz bandwidth.

\subsection{Near-field Localization with Hologram Algorithm}\label{ssec:hologram primer}

Like most recent RFID localization systems, \sys locates a tag under the \textit{near-field} condition, which differs from locating a distant target. Considering the Fraunhofer distance \cite{selvan2017fraunhofer}, a target is at near-field when its distance $d$ from the antenna array meets:

$$d < \frac{2D^2}{\lambda}$$
where $D$ is the aperture of the antenna array, and $\lambda$ is the signal's wavelength. The wavelength of the 915 MHz signal is around 30 cm. When using an antenna array or SAR, the aperture can easily span to 1 m for adequate spatial resolution. $(2D^2)/(\lambda) = (2 \times 1~\text{m}^2)/(0.3~\text{m})$ = 6.7 m and $d < 6.7$ m under most circumstances. Therefore, the response from a tag does not form a plane wave when reaching different elements in the antenna array.

We propose to develop our localization algorithm on top of hologram-based localization framework, which essentially identifies the most likely location as the location estimation, independent of plane wave incidence conditions. 
In the basic hologram algorithm, the theoretical phase $\theta(g_{(i,j)},A_{k},f_{l})$ of a tag at location $g_{(i,j)}$ received by an antenna $A_{k}$ at frequency $f_{l}$ can be written as: 
$$\theta(g_{(i,j)},k,l) =\frac{2\pi f_l}{c}(d_{Tx-Tag}+d_{Tag-Rx})\quad\  (\mathrm{mod}\ 2\pi)$$
where $d_{Tx-Tag}$ and $d_{Tag-Rx}$ are the distance between the tag and the transmitter and receiver, respectively.
For location $g_{(i,j)}$, its likelihood $P(g_{(i,j)})$ of being the tag's true location can be measured by the similarity between empirically received phase $\phi_{k, l}$ from $l$th carrier at $k$th antenna and the theoretically modeled phase $\theta(g_{(i,j)},k,l)$. The hologram algorithm makes the similarity comparison across multiple antennas and frequencies. $P(g_{(i,j)})$ can be written using the following equation:
\begin{equation}\label{eqn: basic hologram}
    P(g_{(i,j)}) = \left|\sum_{l = 1}^{L}\sum_{k = 1}^{K} e^{-j(\phi_{k,l} - \theta(g_{i,j},k,l))}\right|
\end{equation}
Then we can estimation the location of the tag by choosing $(i,j)$ with maximum $P$.

\subsection{Kernel-layer Framework}\label{hologram}
Beyond the basic hologram algorithm \cite{miesen2011holographic}, there are many hologram variants \cite{yang2014tagoram, shangguan2016design,xu2019faho,xu2019adarf}. We find that two key factors determine the performance of hologram-based localization algorithms, namely, kernel and layer:

\nosection{Definition 1.} Kernel is the function that measures the similarity between the received signal and the theoretical signal from one channel (\ie single carrier from a single antenna). For example, the $e^{j(\phi - \theta)}$ in \eqnref{eqn: basic hologram} is a kernel function that measures the phase similarity with an exponential function.

\nosection{Definition 2.} Layer is a function that determines how to combine kernels from multiple channels (\ie multiple carriers from multiple antennas) and obtain the location estimation. For example, the $\sum_{l = 1}^{L}\sum_{k = 1}^{K}$ in \eqnref{eqn: basic hologram} is a layer function.

We introduce a kernel-layer framework that tells us how kernel and layer affect the localization performance.
\figref{fig: kernel layer framework} summarizes our kernel-layer framework, which describes the fundamentals of hologram-based algorithms. This framework can be used following these steps: 

\nosubsection{Model} calculates the theoretical channel information (\eg propagation phase) for each location.

\nosubsection{Measurement} obtains the empirical channel information (\eg propagation phase and RSSI) by interrogating the tags.

\nosubsection{Kernel} function profiles the similarity between the theoretical and empirical channel information.

\nosubsection{Layer} function combines kernel function output from different antennas and frequencies.

\nosubsection{Output} picks the location with the maximum likelihood as the estimated location.

Different kernels and layers can be combined into various near-field localization algorithms. See more examples in \secref{sec: ToF and AoA}.

\begin{figure}[t]
    \begin{centering}
        \includegraphics[width=\linewidth]{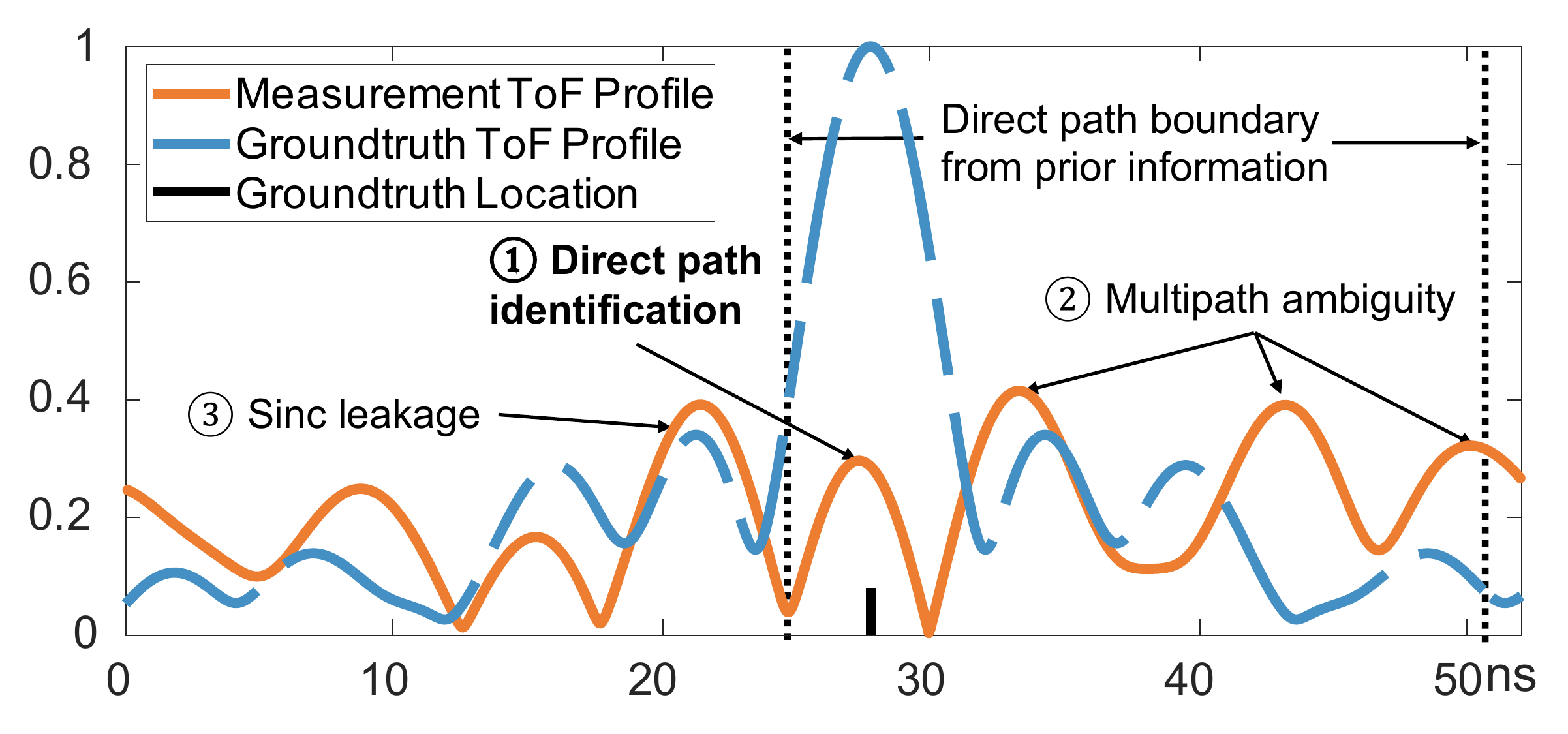} 
    \par\end{centering}
	\vspace{-3mm}
    \caption{\label{fig: find direct path} Direct Path Identification with ROI Information.}
	\vspace{-5mm}
\end{figure}

\subsection{{\sys}'s Kernel and Layer}\label{ssec: sys kernel and layer}
We design our localization algorithm based the kernel-layer framework. When designing {\sys}'s kernel and layer, we want to reduce the impact of multipath for low long-tail error, which can be achieved with the carefully designed kernel, layer, and prior information from the logistic scenario.
{\sys}'s kernel is similar to basic hologram algorithms:
\begin{equation}
    \label{eqn: kernel}
    \text{{\sys}'s kernel: } e^{-j(\phi - \theta)} \nonumber
\end{equation}

\sys has 4 layer functions: ToF estimation layer, direct path identification layer, direct path enhancement layer, and summation layer. These layers work together to suppress the multipath and combat long-tail localization errors.

\nosection{ToF Profile Layer.}
By using the wideband bandwidth captured, this layer computes the time-of-flight profile of the received signal. The computation follows \eqnref{eqn:tof} where $\phi$ is the empirically measured phase value, $f_l$ is the frequency, and $\tau$ is the propagation delay of each path.

\vspace{-2mm}
\begin{equation}
\label{eqn:tof}
\begin{aligned}
\vspace{-3mm}
\text{ToF estimation layer: } S(\tau)  = \sum_{l=0}^{L} e^{-j(\phi_l - 2\pi f_l \tau)}
\end{aligned}
\end{equation}
\vspace{-3mm}

\nosection{Direct Path Identification Layer.}
It is still challenging to identify the direct path in the ToF profile layer in \figref{fig: find direct path} because there are three interfering factors:
\circled{1} If the difference is smaller than 0.75 m, we can only observe one mixed peak in the time-of-flight profile of the received signal. 
\circled{2} If the difference is larger than 0.75 m, there will be ambiguity from multipath at the locations farther from the groundtruth. 
\circled{3} The sample on the frequency domain, which is a sinc function on the time domain, may leak its side lobe and form fake peaks at a nearer location than the groundtruth.
To address these problems, \sys leverages a key observation: prior information. In practical logistic deployment, we can employ the size of the scanning area, the track of tags, \etc to help localization. \sys constructs a layer that leverages this prior information for direct path identification. \figref{fig: find direct path} shows an example of this layer with scanning range [a,b] in meters as prior information, which is common in warehouse deployment. The corresponding algorithm is shown in \algref{alg: direct path identification}. In this example, we first compute the bound of the theoretical propagation time in this range $\tau_a = a/(3\times 10^8)$ and $\tau_b = b/(3\times 10^8)$.
The prior information, $\tau_a$ and $\tau_b$, acts as a filter that eliminates any multipath with a propagation time smaller than $\tau_a$ or larger than $\tau_b$, which helps us identify the right direct path (right peak) rather than nearer one from sinc leakage or farther one from multipath.

\begin{algorithm} 
	\caption{Direct path identification layer} 
	\label{alg: direct path identification} 
	\begin{algorithmic}
		\REQUIRE
		{1. ToF profile: $[S(\tau_1), S(\tau_2), ..., S(\tau_s)]_{1\times s}$}\\
		{2. Prior info: scanning area in meters [a,b]}\\
		{3. Peak threshold: p}
		\ENSURE Direct path distance rough estimation $\tilde{d}_{0}$
		\STATE 1. $\tilde{d}_{0}$ = 0, $\tau_a = \frac{a}{3\times 10^8}$, $\tau_b = \frac{b}{3\times 10^8}$;
		\STATE 2. L = find $\tau_i$ closest to $\tau_a$ in [$\tau_1$, $\tau_2$, ..., $\tau_s$], return index;
		\STATE 3. R = find $\tau_i$ closest to $\tau_b$ in $[\tau_1, \tau_2, ..., \tau_s]$ , return index;
		\STATE 4. $S(\tau) \gets S(\tau)[\text{L}:\text{R}]$;
		\STATE 5. $path \gets S(\tau)[0:\text{end}-1] - S(\tau)[1:\text{end}]$
		\FOR{$i \gets 1$ to $s-1$}{
			\IF{$path[i] > 0$ \& $path[i-1] < 0$ \& $S(i) >$ p}
        	    \STATE $\tilde{d}_{0} = \tau_i\times 3\times 10^8$;
        		\STATE break;
    		\ENDIF 
		}
		\ENDFOR 
	\end{algorithmic} 
\end{algorithm}

\nosection{Direct Path Enhancement Layer.}
{\sys} uses a across-frequency phase redress algorithm to further enhance the signal quality of the direct path signal. 
\sys first identifies potential multipath -- if there are multiple peaks (identified by 2D peak find algorithm \cite{fastpeakfind}) in the basic hologram results, the location estimation is likely affected by the multipath effect. Instead of using the empirically measured phase $\phi$, \sys combines the direct path signal from all frequencies and constructs an enhanced phase $\tilde{\phi}_l$. This process is done by the layer function of \eqnref{eqn: direct path enhancement}. See \secref{sec: time domain beamforming} for detailed mathematical derivation.
\vspace{-4mm}
\begin{equation}
\label{eqn: direct path enhancement}
\begin{aligned}
    \text{Direct path enhancement layer: }
    \tilde{\phi}_l = \angle\sum_{i=1}^{L} e^{j\phi_i}e^{j\frac{2\pi}{c}(f_i - f_l)\tilde{d}_{0}}
    \end{aligned}
\end{equation}

\nosection{Summation Layer.}
The last layer in {\sys} is the summation layer, which combines information from all $L$ frequencies and $K$ antennas and computes the likelihood of the tag position. For every location $g_{(i,j)}$, \sys computes the likelihood $P(g_{(i,j)})$ and choose the position with the highest likelihood as the estimated result.

\vspace{-3mm}
\begin{equation}
\label{eqn: summation layer}
\begin{aligned}
    \text{Summation layer: } &P(g_{(i,j)}) = \left|\sum_{l=0}^L\sum_{k=0}^K 
    e^{-j(\tilde{\phi_{l,k}}-\theta(g_{(i,j)},l,k))}\right|
\end{aligned}
\end{equation}
\vspace{-3mm}

\begin{figure}
	\centering
	\begin{subfigure}{.495\linewidth}
		\includegraphics[width=\linewidth]{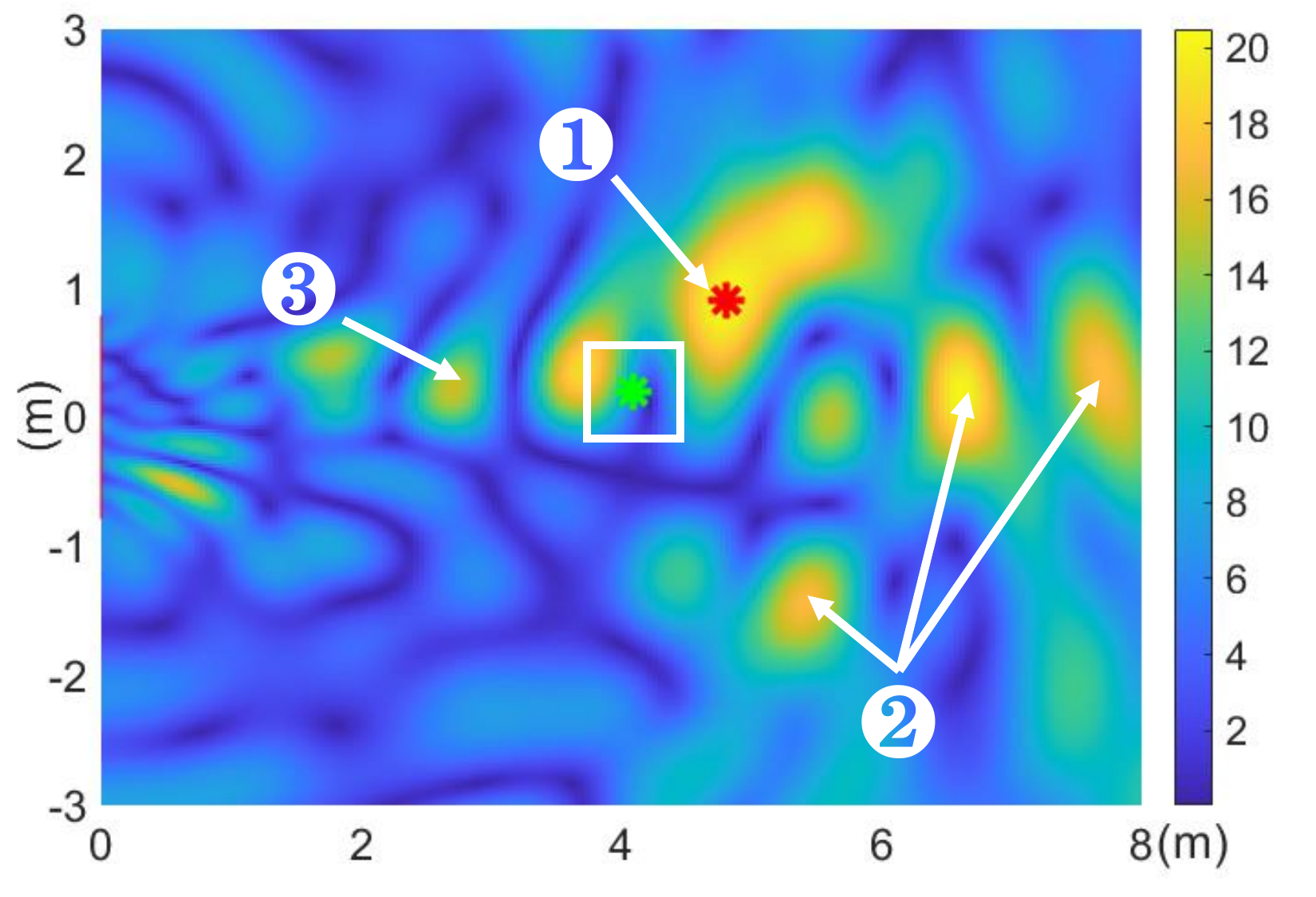}
		\vspace{-5mm}
		\caption{Basic Hologram.}
		\label{fig: basic hologram result}
	\end{subfigure}
	\begin{subfigure}{.495\linewidth}
		\includegraphics[width=\linewidth]{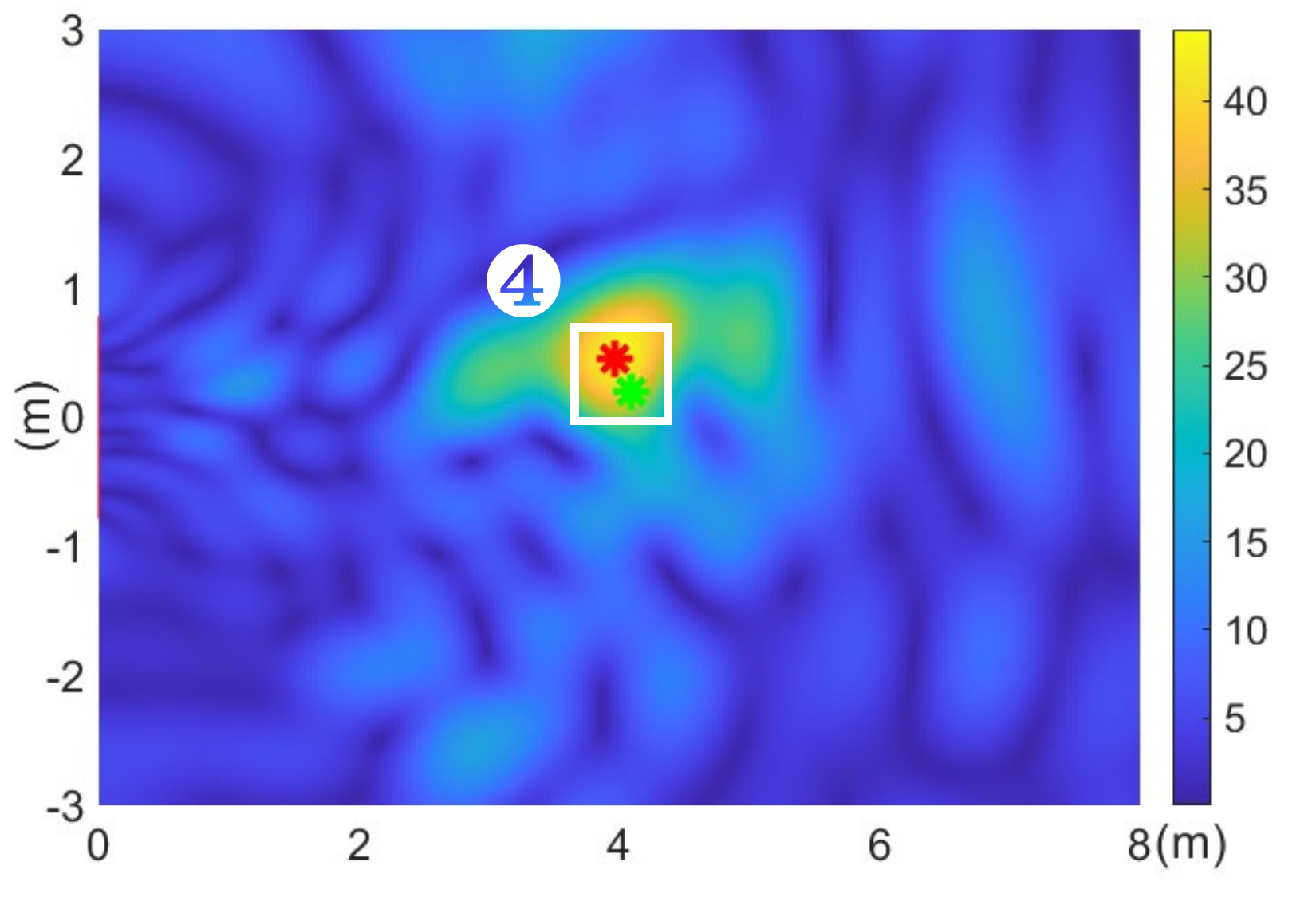}
		\vspace{-5mm}
		\caption{\sys's Algorithm.}
		\label{fig: multipath suppression result}
	\end{subfigure}
\vspace{-6mm}
\caption{\sys can suppress sinc leakage, multipath ambiguity, and enhance direct path for finer resolution compared to basic hologram algorithms in \eqnref{eqn: basic hologram}.}
\vspace{-4mm}
\label{fig: algorithm compare}
\end{figure}
\nosection{Putting Everything Together.}
All above layers and kernel work together as our multipath suppression algorithm.
\figref{fig: algorithm compare} shows an visual example. The heatmaps are the location likelihood with the basic summation layer in \eqnref{eqn: basic hologram} (\figref{fig: basic hologram result}) and with our direct path enhancement algorithm (\figref{fig: multipath suppression result}). The green cross is groundtruth and the red cross is location estimation. 
If we only use the simple summation layer, there are three factors disturbing the localization accuracy. \sys handles them with customized kernel-layer algorithm design.
The peak of location estimation \circled{1} is the superimposed responses from all the paths within distance resolution nearer the direct path. \sys utilizes coherent summation layer with full 200 MHz bandwidth to increase distance resolution to 0.75 m.
The paths with large distance differences from the direct path will generate ambiguity at farther arrival distances as multipath ambiguities \circled{2} or even at nearer distance as sinc leakage \circled{3}.
By using prior information of work range (tags are in different check-in passage with different ranges) to clarify the direct path identification and using direct path enhancement to suppress multipath, we obtain the accurate location estimation \circled{4}.

%% file: 07-impl.tex
\section{Implementation}\label{sec:impl}

\subsection{Active Sniffer}
\nosection{Antenna.} 
We chose a recent variant \cite{zheng2017wideband} of the Foursquare patch antenna \cite{suh2003low}, which is metal-backed and of concentric dual-polarization, as our wideband Tx and Rx antennas for its advantages of small-size, low-cost, and high adaptability to surroundings. The original antenna design is for 1.7\textasciitilde2.7 GHz LTE and we scaled it with HFSS \cite{HFSS} to fit the UHF band 700\textasciitilde1100 MHz. We also attached each Rx antenna to a 915 MHz bandstop filter \cite{bandstopfilter} to suppress the high-power ISM-band leakage from the commercial reader.

\nosection{Array.} 
We built the Rx array through a laser-cutting sheet of aluminum. The mounting holes and SMA clearances on the sheet define a 1$\times$8 linear array with element spacing of 21 cm. We set a notable 31.5 cm gap in the middle for a 2:3 co-prime array configuration \cite{tan2014direction} to suppress the grating lobe.
We hang two Txs 0.4 m lower than the receiver's horizontal array along its geometric bisection. The right one was wideband Tx and the left one was ISM-band Tx.

\nosection{Baseband Processor.}
One of the key implementation challenges towards one-shot inventory is to convert the 31 Gbps I/Q samples from the A/D to the application processor. We developed high throughput baseband with 2 ADRV9009 \cite{mclaurin2018highly, adrv9009} RF chips and an XCKU060 FPGA SoM \cite{ultrascale,ku060som} in charge of 4 receivers over 200 MHz bandwidth for PCIe streaming.

\nosection{Application Processor.}
The host is equipped with a Core-i9 9900 CPU and an RTX 3090 GPU for real-time decoding and CSI acquisition. GPU was used to handle the template matching during the decoding with FFT convolution acceleration and parallelism.
We used Process Explorer \cite{resource} to measure resource utilization and report the results in \tabref{tab: resource}. The decoder is developed with C++/Eigen except that the most compute-intensive part, \ie the full packet matching algorithm, is implemented on GPU with CUFFT \cite{cufft_lib}.

\begin{table}[H]
\centering
\resizebox{\columnwidth}{!}{%
\begin{tabular}{llll}
\toprule
CPU (Utilization) & GPU (Utilization) & I/O Bandwidth & Memory \\
\midrule
Core-i9 9900 (16.1\%) & RTX 3090 (38.0\%) & 520.1 MBps             & 4.1 GB    \\
\bottomrule
\end{tabular}%
}
\vspace{-2mm}
\caption{Hardware Resource Utilization.}
\vspace{-2mm}
\label{tab: resource}
\end{table}

\subsection{RFID Tags}
In order to ensure compatibility and low-cost, we used a commercial RFID IC Impinj Monza-M4A \cite{tagchip} and implemented a bandwidth extension technique \cite{deavours2009analysis} to redesign the metal inlay (antenna) on 80 $\times$ 80 mm single-sided PCB. The CAD of the RFID antenna is shown at the top left of \figref{fig: exp setup} and its direction gain (similar to dipole antenna) is shown in \figref{fig: orientation setup}. It works on 700\textasciitilde1000 MHz, whose copper geometry can be transferred to flexible inlay for massive production.

%% file: 08-eval.tex
\begin{figure}[t]
    \centering
    \includegraphics[width=\linewidth]{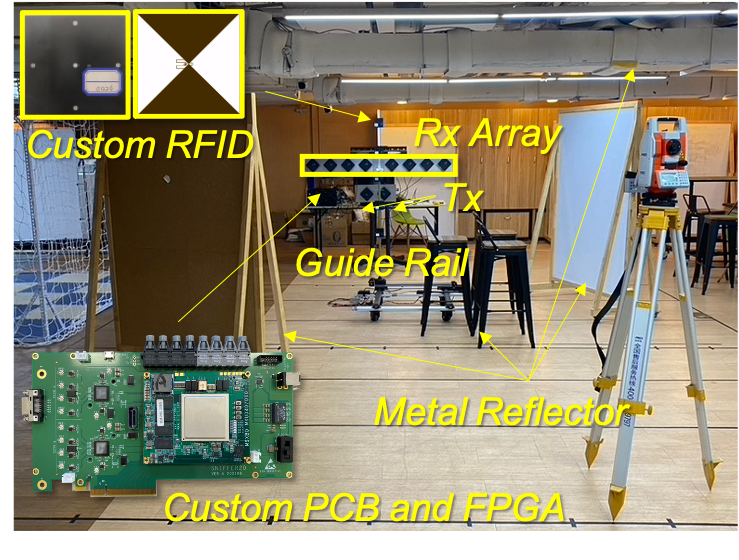}
    \vspace{-6mm}   
    \caption{Experimental setup for evaluating performance. Five tags are mounted on the rail and \textasciitilde20k tag responses are collected in 384 locations.}
    \vspace{-3mm}
    \label{fig: exp setup}
\end{figure}

\section{Evaluation}\label{sec:eval}

\subsection{Experimental Setup}

\nosection{Testing Environment.} 
We evaluate \sys in an office with multiple reflectors (\eg metal furniture, low ceilings, and walls). The evaluation range is the area of 6 $\times$ 3.2 m ahead of the antenna. We divide the evaluation space into 20 cm grids and use guide rails to move the tags. All the tags are facing the array. The dataset containing about 20k wideband RFID channel information measurements at 384 locations is open-sourced at \cite{open_source}. The setup is shown in \figref{fig: exp setup}.

\nosection{Location Groundtruth.} Groundtruth is measured from a total station theodolite (TST) \cite{totalstation} with a 2 mm/2$''$ accuracy.

\nosection{Frequency Band Configuration of Active Sniffer.} 
We use the band of 787\textasciitilde987 MHz and avoid selecting carriers in ISM band 902\textasciitilde928 MHz. The carriers are almost evenly selected with spacing of 11.1 MHz\footnote{The frequency set of carriers is \{787.1, 798.2, 809.3, 820.4, 831.5, 842.6, 853.7, 864.8, 875.9, 887.0, 898.1, 942.5, 953.6, 964.7, 975.8, 986.9 MHz\}.}. The spectrum analyzer shows the inter-modulation distortion of carriers is very little.

\nosection{ISM-band Reader.} 
We use an Impinj R700 \cite{r700} as the ISM-band reader, which is configured on ``Radio Mode 142'' (Miller-4 coding and BLF of 256 kHz) and a single linear-polarized antenna aligned with the wideband Tx.
We empirically pick this mode since it balances throughput and range. Other coding methods and BLF can also be adopted with few modifications to our system. 

\subsection{Throughput in One-shot Localization}\label{ssec: throughput}
\begin{figure}[t]
    \centering
\includegraphics[width=.99\linewidth]{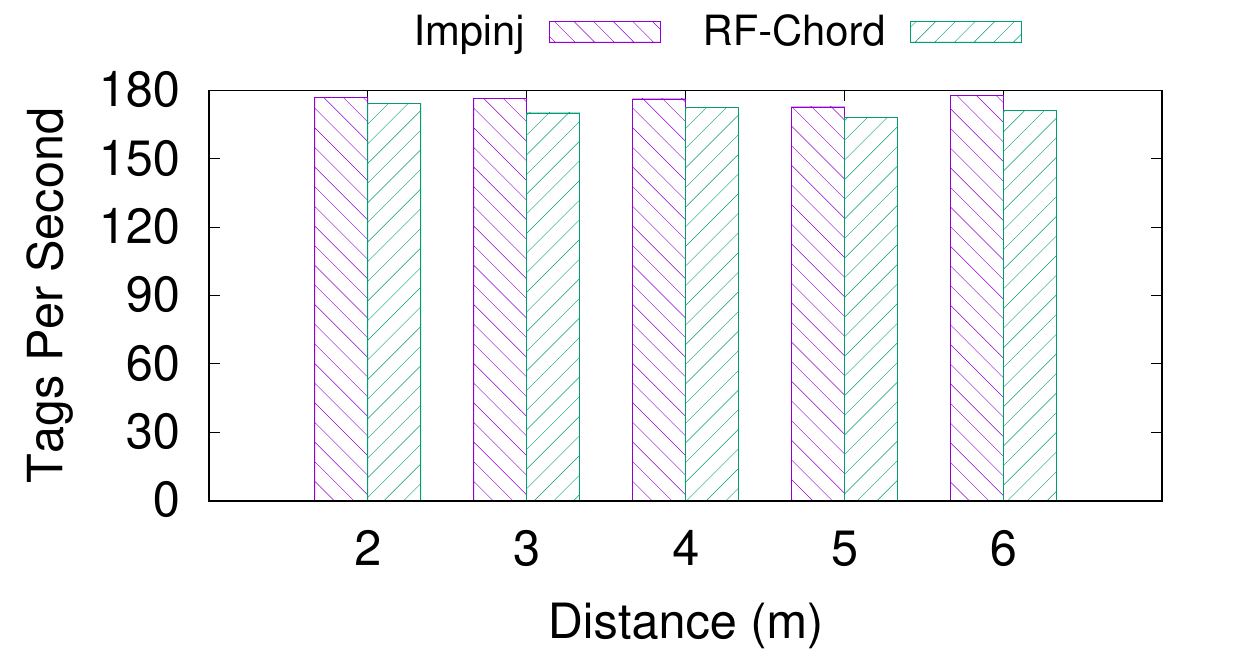}
    \vspace{-1mm}
    \caption{Throughput across distances. \sys can localize around 180 tags/s with -15 dBm emission power.}
    \vspace{-1mm}
    \label{fig:throughput}
\end{figure}

\begin{figure}[t]
    \centering
\includegraphics[width=.99\linewidth]{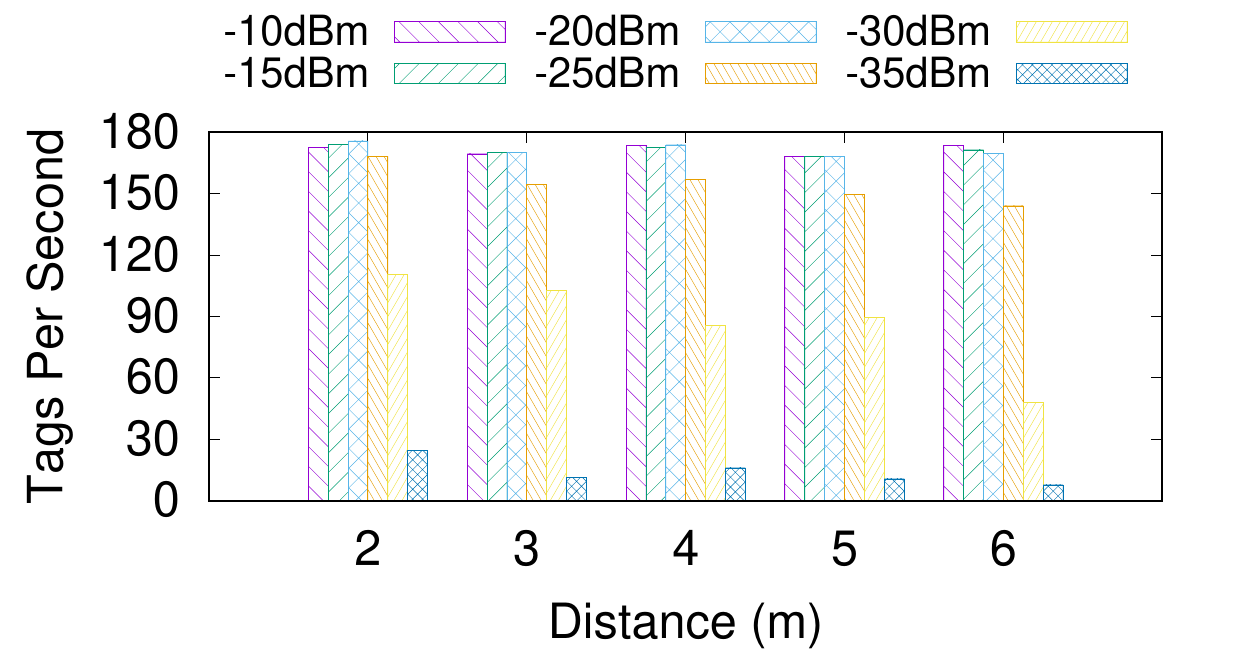}
    \vspace{-1mm}
    \caption{Throughput across distances with different emission power. The performance of \sys is stable with above -25 dBm emission power.}
    \vspace{-3mm}
    \label{fig:throughput_power}
\end{figure}

\figref{fig:throughput} shows \sys's throughput at different distance. \sys can read and localize \textasciitilde180 tags per second (97\% of the tags read by an Impinj reader) at up to 6 m.
\sys is 1000$\times$ faster compared to previous sniffer-based wideband systems with frequency-hopping. For instance, RFind \cite{ma2017minding} needs 6.4 seconds to localize one tag.
We also evaluate \sys's throughput across emission power. \figref{fig:throughput_power} shows that \sys's throughput decreases when we reduce its emission peak power from -15 dBm to -35 dBm. It works fine with an emission power above -25 dBm.

\begin{figure*}[t]
	\centering
	\begin{subfigure}{.33\linewidth}
        \includegraphics[width= \linewidth]{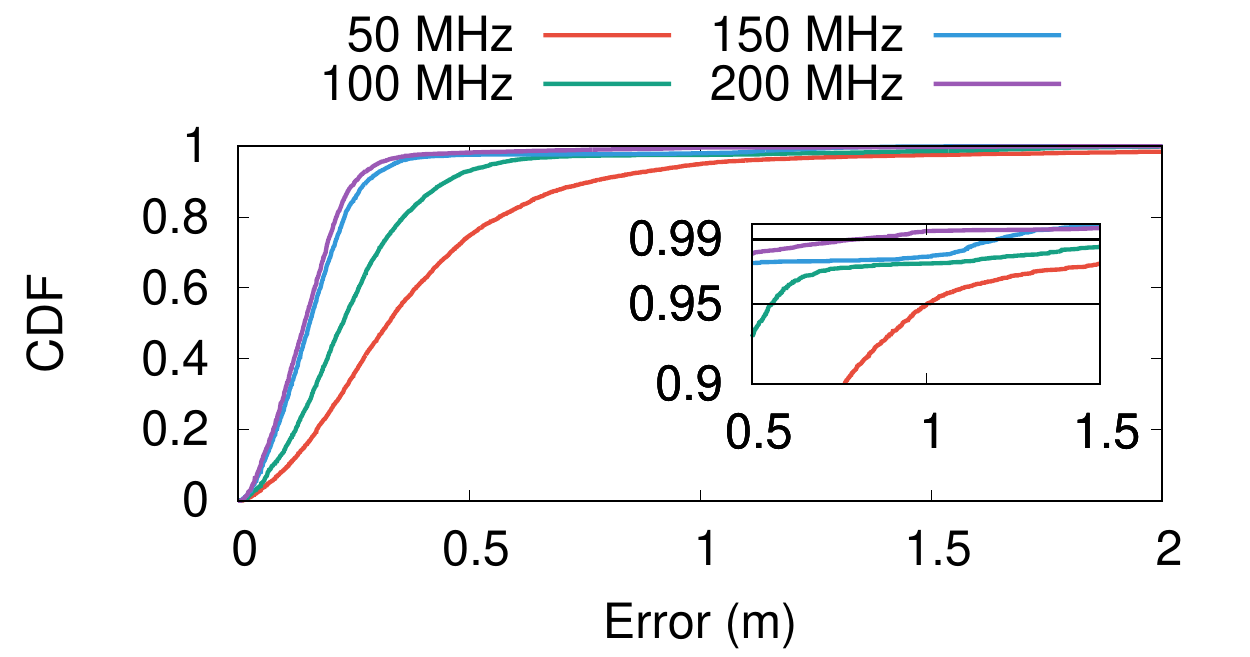}
        \vspace{-5mm}
        \caption{Bandwidth.}
        \vspace{-3mm}
        \label{fig: bandwidth}
	\end{subfigure}
	\begin{subfigure}{.33\linewidth}
		\includegraphics[width=\linewidth]{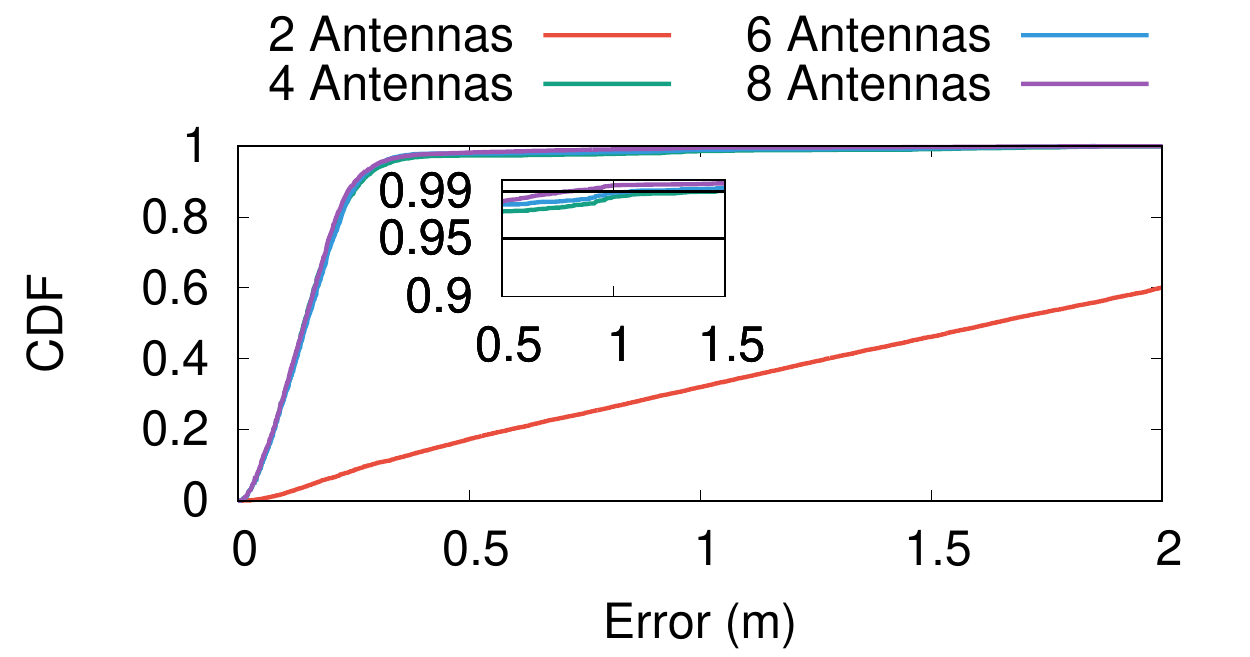}
		\vspace{-5mm}
		\caption{Number of Antennas.}
		\vspace{-3mm}
		\label{fig: antenna number}
	\end{subfigure}
	\begin{subfigure}{.33\linewidth}
        \includegraphics[width=\linewidth]{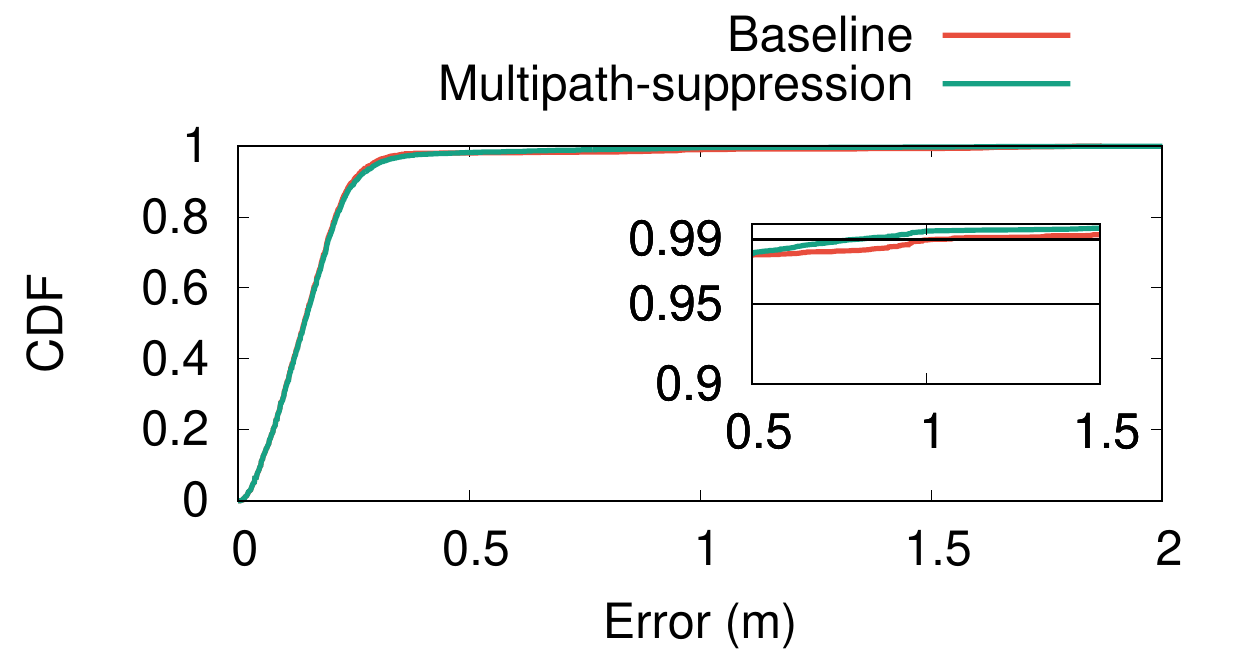}
        \vspace{-5mm}
        \caption{Algorithm.}
        \vspace{-3mm}
        \label{fig:overall}
	\end{subfigure}
	\vspace{-3mm}
	\caption{{\sys}'s localization errors with different bandwidths, antenna numbers, and algorithms.}
	\vspace{-3mm}
	\label{fig: performance}
\end{figure*}
\begin{figure*}[t]
	\centering
	\begin{subfigure}{.20\linewidth}
	    \centering
		\includegraphics[width=\linewidth]{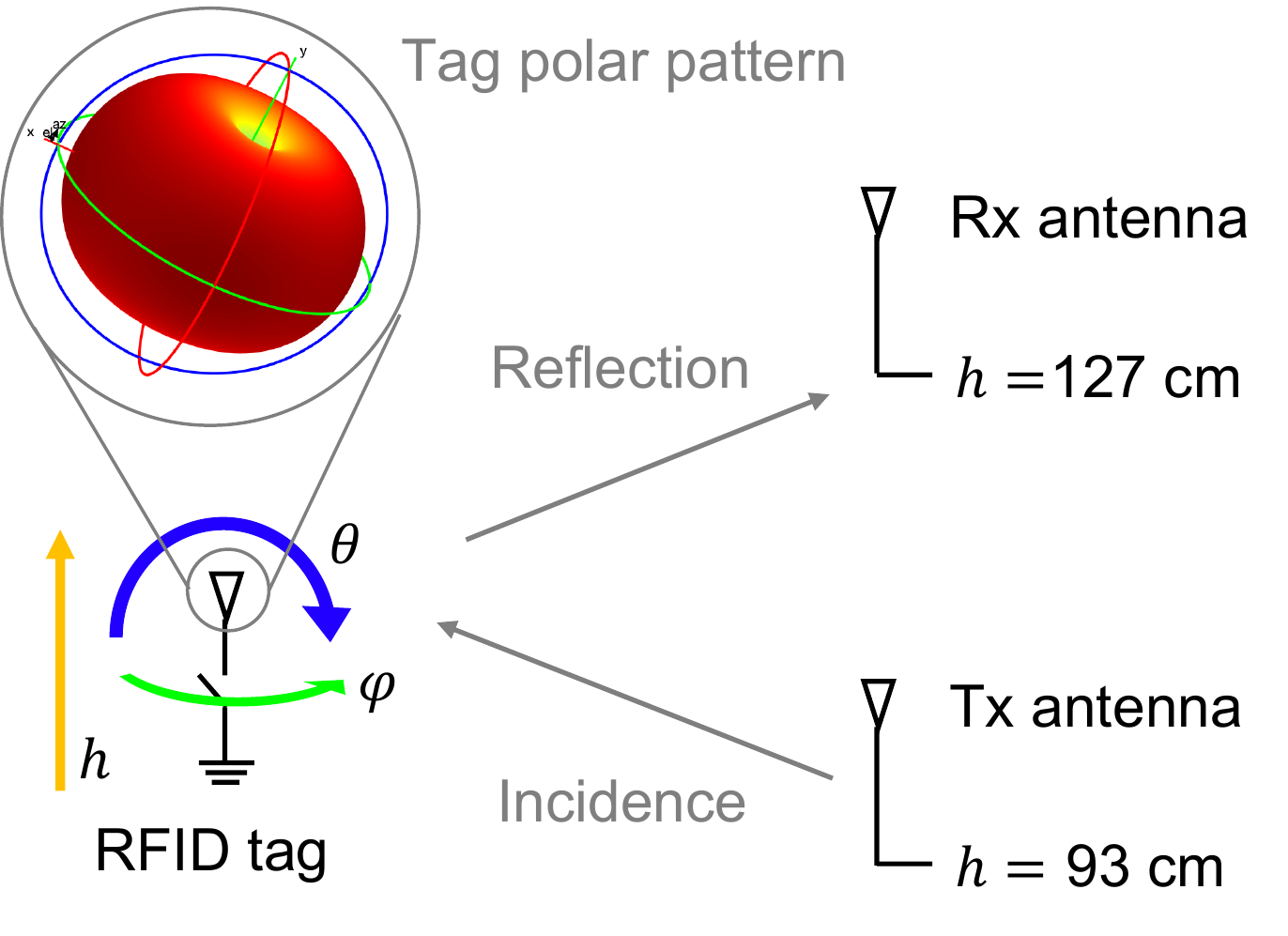}
		\vspace{-5mm}
		\caption{Orientation Setup.}
		\vspace{1mm}
		\label{fig: orientation setup}
	\end{subfigure}
	\begin{subfigure}{.25\linewidth}
	    \centering
		\includegraphics[width=\linewidth]{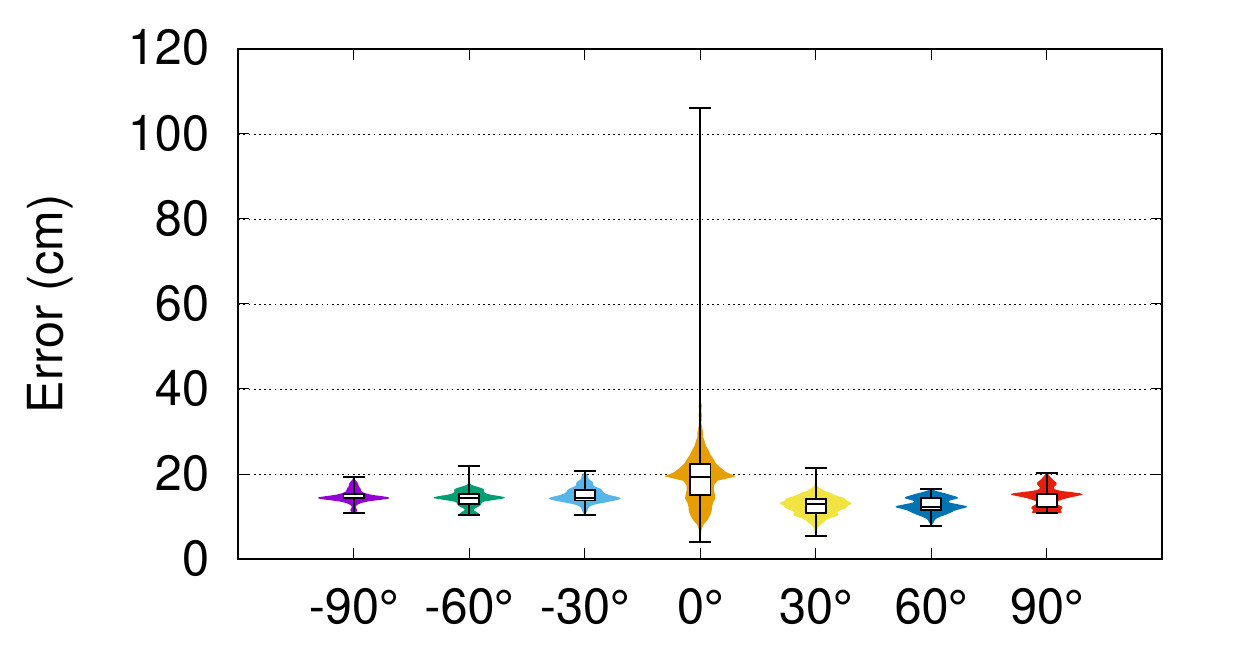}
		\vspace{-2.5mm}
		\caption{Pitch/Roll Angle ($\theta$).}
		\vspace{1mm}
		\label{fig: orientation pitch}
	\end{subfigure}
	\begin{subfigure}{.25\linewidth}
	    \centering
		\includegraphics[width=\linewidth]{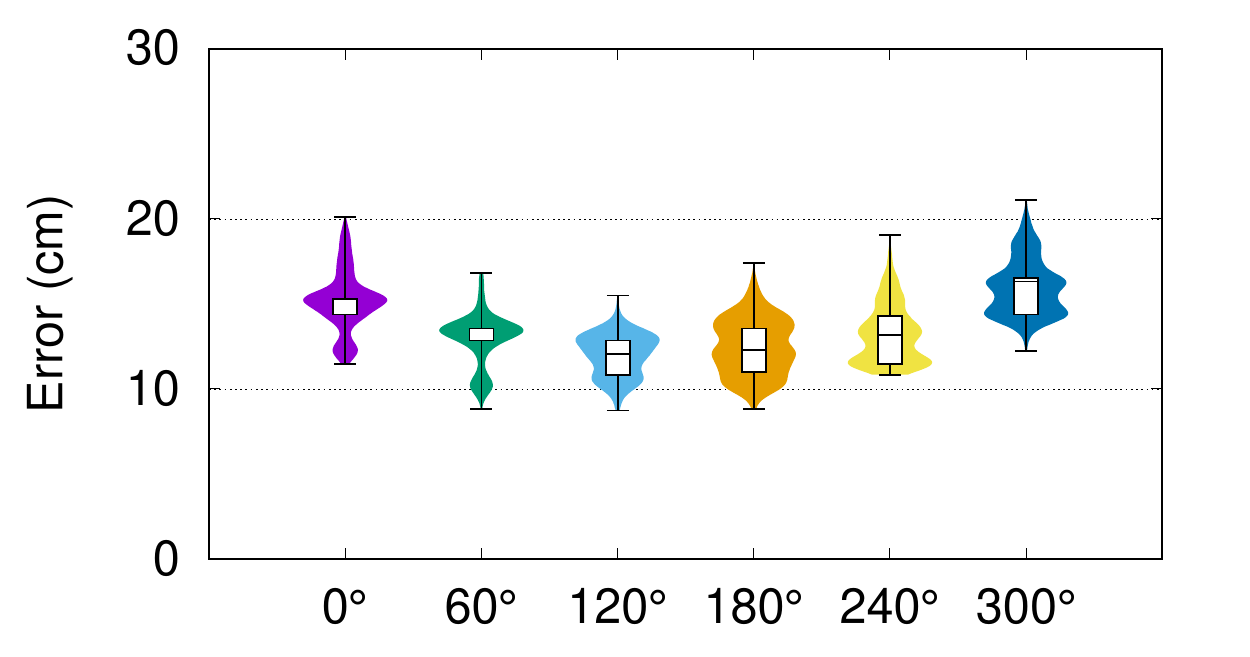}
		\vspace{-2.5mm}
		\caption{Yaw Angle ($\phi$).}
		\vspace{1mm}
		\label{fig: orientation yaw}
	\end{subfigure}
	\begin{subfigure}{.25\linewidth}
	    \centering
		\includegraphics[width=\linewidth]{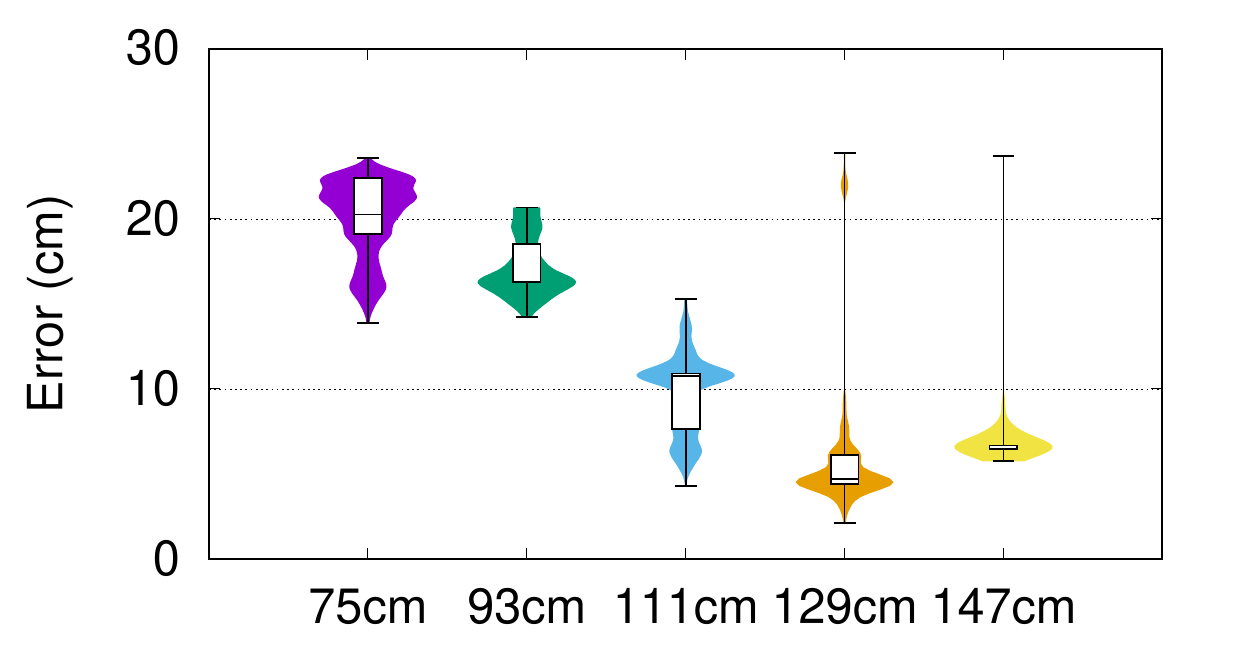}
		\vspace{-2.5mm}
		\caption{Height ($h$).}
		\vspace{1mm}
		\label{fig: orientation height}
	\end{subfigure}
	\vspace{-3mm}
	\caption{Microbenchmarks with different tag orientations and heights related to the antennas.}
	\vspace{-4mm}
	\label{fig:orientation microbenchmark}
\end{figure*}

\subsection{Localization Performance}
\sys utilizes large bandwidth, multiple antennas, and the multipath-suppression algorithm to realize one-shot and high-reliability localization. We conduct microbenchmarks to evaluate how physical resources (frequency and spatial domain), algorithms, and orientation influence the localization.

\nosection{Bandwidth.}
We evaluate the localization performance with 8 antennas and different bandwidths. \figref{fig: bandwidth} shows 99th localization errors are 2.398 m, 1.646 m, 1.203 m and 0.786 m with 50 MHz, 100 MHz, 150 MHz and 200 MHz bandwidths. The median errors are 0.325 m, 0.227 m, 0.155 m, and 0.144 m, separately.
The results show increasing bandwidth, thus increasing the time resolution, can not only improve the median performance but also reduce the long-tail error. Even when the median performance is close to the upper limit (150 MHz v.s. 200 MHz), the long-tail errors can still be reduced by increasing bandwidth.

\nosection{Number of Antennas.}
We evaluate \sys's localization performance with 200 MHz bandwidth and different numbers of antennas (thus different array apertures). \figref{fig: antenna number} shows \sys's 99th localization errors are 4.513 m, 1.467 m, 1.081 m and 0.786 m when 2, 4, 6 and 8 antennas are used. The performance of the 4, 6, and 8 antennas is very similar on median errors (about 0.14 m). However, their long-tail errors are significantly different. The results show increasing the number of antennas (from 2 to 8) can always improve long-tail performance.
Increasing the number of antennas/apertures strengthens the system's immunity to interference in specific directions and improves the angular resolution for localization.

\nosection{Algorithms.}
We take basic hologram (\eqnref{eqn: basic hologram}) as the baseline algorithm and evaluate our multipath-suppression algorithm with 8 antennas and 200 MHz bandwidth.
\figref{fig:overall} shows that 99th localization errors of baseline and {\sys} are 1.018 m and 0.786 m respectively. The median errors of baseline and {\sys} are 0.143 m and 0.144 m, respectively. The algorithm effort can improve long-tail performance by handling more corner cases, but hard to improve median performance. Physical resources (\ie the bandwidth and the antenna array aperture) fundamentally limit the algorithm's performance, and the long-tail improvement from the algorithm is primarily attributed to the introduction of prior information -- it provides an appropriate carrier for making use of prior information.

\begin{figure*}[]
\begin{minipage}[t]{.38\linewidth}
\centering
\includegraphics[width=\linewidth]{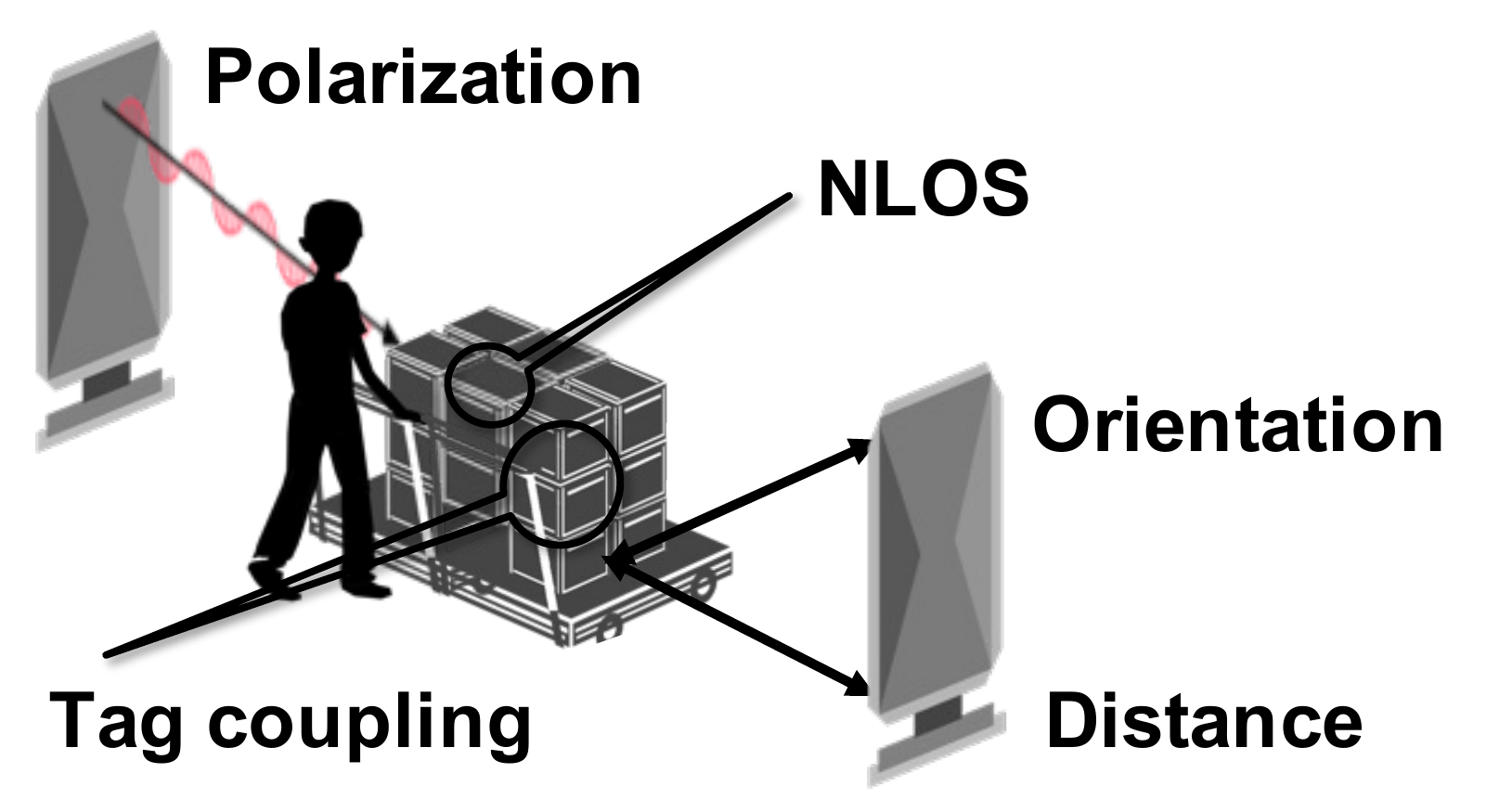}
    \vspace{-3mm}
    \caption{The Factors Affecting Signal Quality.}
    \vspace{-1mm}
    \label{fig: signal quality}
\end{minipage} 
\hfill
\begin{minipage}[t]{.3\linewidth}
\centering
    \includegraphics[width= \linewidth]{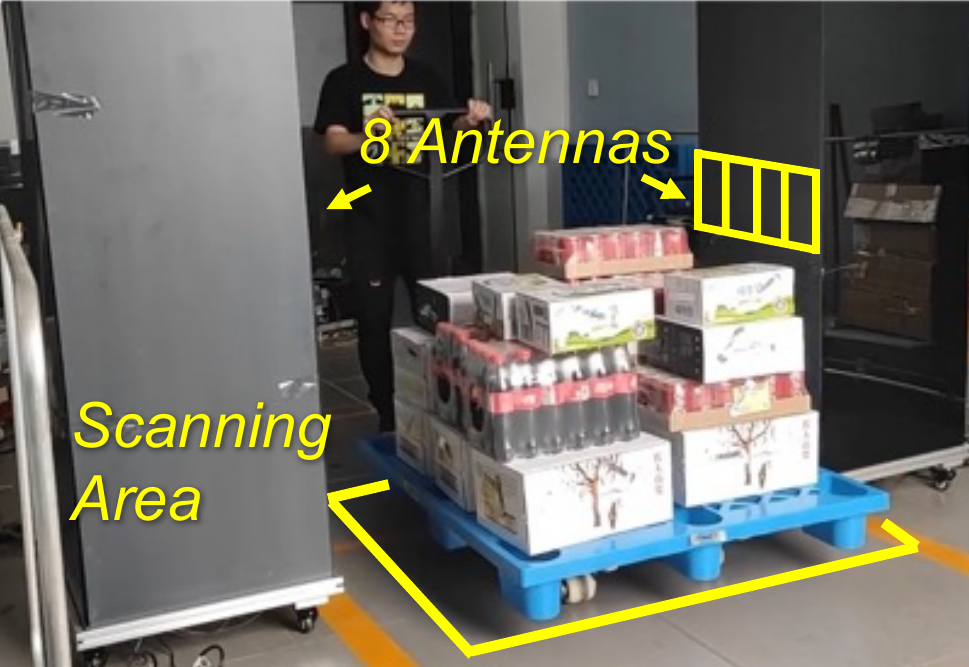}
    \vspace{-3mm}
    \caption{Warehouse Dock Door.}
        \vspace{-1mm}
    \label{fig: warehouse setup}
\end{minipage}
\hfill
\begin{minipage}[t]{.3\linewidth}
		\includegraphics[width=\linewidth]{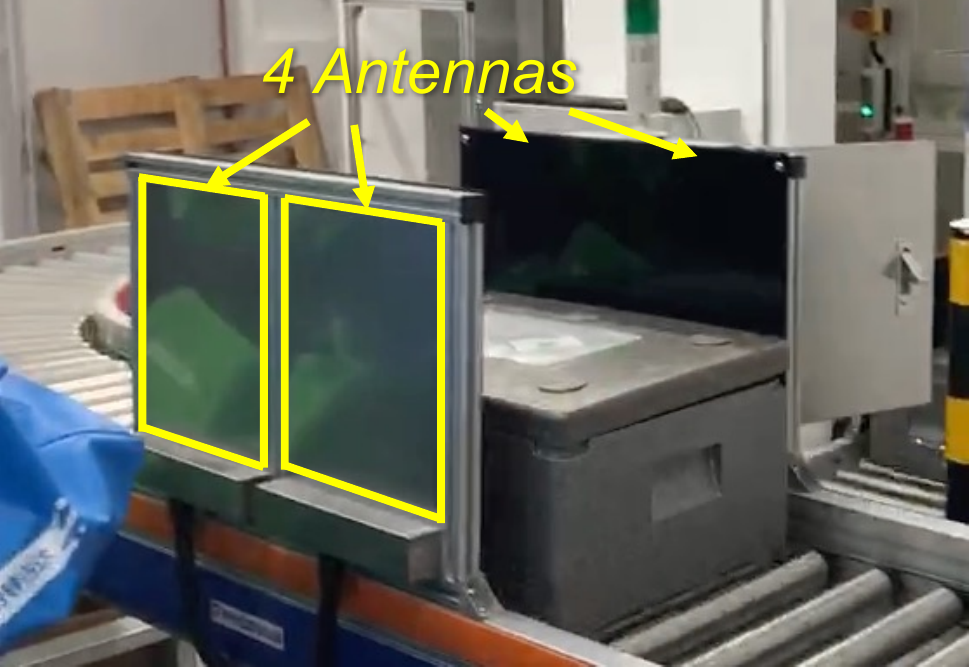}
		\vspace{-3mm}
		\caption{Food Delivery Store.}
		\vspace{-1mm}
		\label{fig: food delivery setup}
\end{minipage}
\end{figure*}

\nosection{Orientation.}\label{ssec: orientation}
In practice, the orientation of tags will influence the link angle and polarization, thus introducing SINR and phase changes. We evaluate how orientation influences the localization error.
We set the target tag at a 1-m fixed distance to the antenna array to eliminate the influence of the multipath effect.
Then, we change the pitch angle $\theta$ (as same as the roll angle due to the symmetry), yaw angle $\phi$, and height of the tags as shown in \figref{fig: orientation setup} for orientation microbenchmark: 

\nosubsection{Pitch/Roll Angle.}
In \figref{fig: orientation pitch}, we keep $\phi = 0\degree$ and $h = 111$ cm (at the center between Tx and Rx).
The worst performance occurs when the pole of the antenna points to the rx, which rarely happens in practical deployments (to be discussed in \secref{ssec: deployment detail}). It is difficult to read tags due to the low SINR, and even if successful, the long-tail error will be more than 1 m. 

\nosubsection{Yaw Angle.}
In \figref{fig: orientation yaw}, we keep $\theta = 0\degree$, $h = 111$ cm and change $\phi$ from $0\degree$ to $300\degree$. The errors at different yaw angles are similar because the directional gain across $\phi$ is symmetrical. The results show that the yaw angle does not affect long-tail localization error (bounded within 30 cm).

\nosubsection{Height.}
In \figref{fig: orientation height}, we keep $\theta = 0 \degree, \phi = 0\degree$ and move the tag from 75 cm to 147 cm. The long-tail errors do not change much across different heights, which shows that height is not the key factor affecting long-tail errors.

\section{Practical Deployment}
\label{sec: practical deployment}

\subsection{Deployment Constraints}\label{ssec: deployment detail}
We summarize the practical factors that influence SINR in \figref{fig: signal quality} and introduce the constraints in real-world logistic scenarios. We also explain how we avoid or utilize them for high-reliability localization.

\nosection{Orientation.}
The localization error may be significant if the pitch angle of the tag is closed to 90$\degree$ according to \secref{ssec: orientation}.
In the deployment shown as \figref{fig: warehouse setup}, the orientation of tags may not be uniform but unlikely to be completely disordered. All the tags are attached to the sides of boxes or crates and then stacked on the pallet. The chaos of stacking and the movement of the pallet may cause yaw angle ($\phi$) change but not cause much pitch/roll angle ($\theta$) change, which only introduces negligible localization errors according to \figref{fig: orientation yaw}.

\nosection{Polarization.}
We set the tags and sniffer antennas all vertically polarized, so horizontal tags can not be read. Similar to orientation, no tag will be misplaced in our scene because the pallet stack constraints the crate direction.

\nosection{NLOS and Tag Coupling.}
We also stipulate that all the tags should be in the line of sight from one side dock door, which means stacking at most two-column crates on the pallet. It is because the performance of UHF RFID will decrease rapidly with nearby water \cite{aroor2007evaluation}. This rule excludes severe NLOS occlusion/reflection and severe tag coupling. Most of the pallets in our scenes naturally meet this requirement, and in rare cases, we need to waste some space.

\begin{table}[]
\centering
\resizebox{0.9\linewidth}{!}{%
\begin{tabular}{lll}
\toprule
 & {Miss Reading Rate} & Cross Reading Rate \\ 
\midrule
{Warehouse} & {0}      & 0.0025\%  \\
{Food delivery} & {0} & 0.0154\%  \\
\bottomrule
\end{tabular}%
}
\vspace{-1mm}
\caption{\label{tab:deployment1}Reliability performance in practical deployments.}
\vspace{-3mm}
\label{tab:deployment}
\end{table}
\subsection{Real World Deployment}\label{ssec: real world}
We deployed the full-fledged \sys (\ie 200 MHz bandwidth and 8 antennas) in the warehouse dock door and lightweight \sys (\ie 200 MHz bandwidth and 4 antennas) in the fresh food delivery store according to cost and scene conditions for operational evaluation.

\nosection{Warehouse Deployment.} 
We deploy \sys in a warehouse to understand its performance in logistic check in and out. \sys is installed in the dock door of the warehouse as shown in \figref{fig: warehouse setup}. This warehouse's goal is to distribute a large amount of food and daily necessities supplied by the upstream warehouse to city delivery stations. The crates are various and packaged without unified standards. Ideally, \sys should report all the tags inside of the scanning area and not report any tags outside of the scanning area. 
Our previous deployment experiments in the same scenario show that commercial read-or-not solution Impinj xSpan \cite{impinj_xspan} has \textasciitilde 6\% miss-reading rate and \textasciitilde 2\% cross-reading rate in the similar scene.
We attached over 10,000 tags to various items, mainly plastic crates, also including water bottles, cans, milk boxes, rice, \etc \revision{The scanning area is 2 $\times$ 1 m between the two poles of the dock door (as ROI) and the user walks through the aisle with about 50\textasciitilde100 tags on a trailer in 1 to 4 seconds.} According to \tabref{tab:deployment}, \sys is able to identify tags inside of scanning area with a 100\% accuracy (perfectly no miss reading) and 0.0025\% cross reading.
Therefore, \sys can provide sufficient localization accuracy in the warehouse deployment, which significantly outperforms the state-of-the-art commercial solutions.

\nosection{Fresh Food Deliver Store Deployment.}
As shown in \figref{fig: food delivery setup}, we also deploy lightweight \sys in a fresh food delivery store where fresh food is packaged into a container and transported via a moving belt. Once the RFID tag on the container is scanned, the delivery personnel will be allocated to pick it up. \sys needs to ensure all the containers on the moving belt are scanned and do not scan any tag outside of the moving belt. \tabref{tab:deployment} shows that the miss-reading rate of \sys is 0\%, and cross-reading rate is 0.0154\%. Therefore, \sys can achieve sufficient accuracy in the fresh food delivery store deployment.

%% file: 09.1-discuss.tex
\section{Discussion}\label{sec: discussion}

\nosection{Polarization Mismatch.} 
In our scenarios, the work pipeline guarantees the polarization match. However, in more general scenarios, the polarization may be mismatched when the orientation of tags is disordered. The conventional solution is to use circular polarization antenna \cite{impinj_far} or dual-polarization switching \cite{impinj_xspan}. \sys can be adapted to them conveniently because its wideband four point antenna is inherently dual-polarized. We can plug a polarization switch into each sniffer antenna, which acts synchronously and does not influence throughput and range performance.

\nosection{Blind Spots.}
\sys is free of cross reading, and therefore it can use high transmission power and sensitivity ISM-band reader for achieving nearly zero miss-reading rate. 
However, miss reading still threatens reliability in certain complex environments. It can be mitigated by switching between antennas or beam patterns \cite{bocanegra2020rfgo, wang2019pushing}. As our Tx is synthesizing multiple tones, it is feasible to add a Tx beamforming array for blind spot suppression.

\nosection{Integration with Robots.}
In recent years, logistics robots (\eg automated guided vehicle (AGV) \cite{AGV}, automated storage and retrieval systems (ASRS) \cite{ASRS}, and autonomous mobile robots (AMR) \cite{AMR}) have been developed rapidly to reduce the movements and operations of sorters and improve efficiency. These robots still need to cooperate with a label identification system (\eg barcode or QR code). RF-Chord has the potential to replace such system and cooperate with logistics robots to achieve more efficient automation.

\nosection{Cost.}
The ultimate goal of deploying RFID is to reduce manual labor and error while improving efficiency, which requires careful cost accounting. 
We emphasize that although baseband chips and RF circuits will increase the cost of readers to thousands of dollars, the main cost of RFID-based logistics still comes from RFID tags.
Considering a medium warehouse with 10k packages delivered every day, the annual cost of tags is approximately \$0.1 $\times$ 10,000 $\times$ 365 = \$365,000.
Our strategy is not modifying the tag chip because most of the manufacturing cost comes from the chip and the assembly process \cite{swamy2020manufacturing}. Therefore, the wideband tag we designed maintains almost the exact cost as current commercial tags when in massive manufacturing.

%% file: 03-related.tex
\section{Related Work}
\nosection{Narrowband Localization.}
There are three main localization approaches to boost accuracy even with the limited time resolution of narrow ISM bandwidth:
The first approach is to improve spatial resolution by SAR. 
Tagoram \cite{yang2014tagoram} uses the motion of tags to build multiple virtual antennas, while Mobitagbot \cite{shangguan2016design} exploits antenna motion. The hologram algorithms in these two systems inspired the kernel-layer framework in our paper. Other hologram algorithm variants \cite{xu2019adarf,xu2019faho,zhang2020rf} can also be viewed as different combinations of kernels and layers.
However, the assumption of free antennas or tags mobility and lengthy startup time for tracking do not fit the logistic network.
The second approach is to acquire prior information by reference tag.
PinIt \cite{wang2013dude} exploits a dense grid of reference tags and determines the nearest reference tag for NLOS localization by dynamic time wrapping. However, reference tags share time slots, which influences the throughput and scalability.
The third approach is to increase the number of links by tag array.
Attaching more tags to the target can increase the number of links and improve localization performance.
Tagyro \cite{wei2016gyro}, and RF-Dial \cite{bu2018rf} utilize the phase difference of the tag array to solve orientation ambiguity and improve localization performance.
Trio \cite{ding2018trio} models the equivalent circuits of coupled tag and uses the tag interference for refined localization. 
Liu et al. \cite{shangguan2015relative} uses spatial-temporal phase profiling for relative RFID localization. 
These tag array based localization approaches are accurate but may be error-prone in a complex environment.
Unlike these proposals, \sys is a sniffer-based wideband localization system that improves time resolution for fundamental performance enhancement.

\nosection{Wideband Localization.}
Wideband RFID localization has been proposed to overcome the time resolution limitation.
RFind \cite{ma2017minding} uses a low-power sniffer antenna by frequency hopping to collect the narrow sample channel state information across 220 MHz. 
Turbotrack \cite{luo20193d} develops an OFDM-based one-shot wideband channel estimation approach and a Bayesian space-time super-resolution algorithm to achieve fine-grained localization. 
However, these systems need multiple shots in the channel estimation or the algorithm to converge for fine location estimation, thus very slow startup for localization or tracking.
Modifying tags to work on other frequencies (\eg Wi-Fi \cite{kellogg2014wi}, millimeter-wave \cite{adeyeye2019miniaturized}, UWB \cite{arnitz2009multifrequency, decarli2015passive}) or cross-frequency based approaches (\eg communicate with Wi-Fi \cite{an2018cross}, communicate at 1.4\textasciitilde2.4 GHz \cite{ma2014accurate}) are also expected as the solutions for both finer localization and higher throughput, but their tags are not ready for massive manufacturing at low cost due to the complicated RF frontend and control circuits. 
Inspired by these works, \sys develops a multisine waveform to realize one-shot localization without modifying the commercial tag chips, resulting in high accuracy with no throughput loss or cost increase.

\nosection{RFID Reader.}
Commercial RFID readers \cite{r700,r420,arl9900} have heavily optimized RF analog frontend, decoder, and protocol stack but do not support real-time tag critical information (\ie EPC ID, timestamp) retrieval. 
There are a series of open-source RFID reader systems.
Buettner et al. implemented EPC Gen II downlink stack \cite{buettner2008empirical} and the full functional reader \cite{buettner2011software}, respectively.
Kimionis et al. implemented a GNU radio-based reader, which supported OOK and noncoherent FSK \cite{kimionis2012design}.
However, their energy and edge detection algorithms are too simple to decode applicable code (\eg miller-4 coding).
A recent reader designed by Kragas et al. \cite{kargas2015fully} is featured by coherent detection and initial duration deviation search but only supports simple FM0 encoding. 
There are other research projects featured by multisine waveform \cite{boaventura2017design}, parallel sensing support \cite{wang2021toward}, and active transmit leakage cancellation \cite{keehr2018low}. However, they only focus on specific optimization and do not provide source code. 
In a nutshell, no out-of-box reader design meets our requirements of high throughput and low decoding threshold, so we develop a wideband reader with a customized RF frontend and decoder while reusing the MAC layer of the commercial reader for slot arrangement and collision handling. It supports our wideband localization with high efficiency, sensitivity, and compatibility.

%% file: 09-concl.tex
\section{Conclusion}\label{sec:concl}
We illustrate the three key requirements in reliability, throughput, and range to meet the industry-grade standard of the logistic network, and present \sys, the first RFID system that considers all these factors from wideband signal and baseband processing to localization algorithm framework development. We believe our real-world empirical results demonstrate that \sys paves the way for the practical hardware-software methodological solution of RFID localization-based logistic network and makes an important step towards large-scale operational deployment. 

%% file: 11-ack.tex
\section*{Acknowledgments}
We are grateful to the reviewers for their constructive critique, and our shepherd, Vikram Iyer in particular, for his valuable comments, all of which have helped us greatly improve this paper. We also thank Xieyang Xu and Weicheng Wang for providing an early implementation version of the work. We are grateful to Yunfei Ma for the thoughtful suggestions based on the early version of the work.
This work is supported in part by National Key Research and Development Plan, China (Grant No. 2020YFB1710900), National Natural Science Foundation of China (Grant No. 62022005, 62272010, and 62061146001) and Alibaba Innovative Research. Chenren Xu and Shunmin Zhu are the corresponding authors.

%% file: 10-proofs.tex
\appendix
\section{FCC Compliance} \label{sec:FCC}
\sys adopts a 200 MHz bandwidth in the UHF band, much wider than the  902\textasciitilde928 MHz ISM band. We need to reduce the power of the signal emitted in the licensed band to follow the FCC regulation \cite{FCCUnlicense}. Similar operations exist in other systems, such as RFind \cite{ma2017minding}.
RFind adopts a duty-cycled single-tone signal with a peak power of -3 dBm and average power of -13.3 dBm. However, due to the throughput requirement of the localization, 
\sys's sniffer should always be ready to localize a tag, which means duty cycling is unacceptable. Therefore, \sys adopts a hard limit of -15 dBm per tone and can be even lower with similar performance.
One may concern that the multiple carrier operation will not be the same as RFind \cite{ma2017minding} since the total bandwidth is larger than the 0.25\% bandwidth limitation in FCC 15.231 (c) \cite{FCC15231}. However, \sys can adopt the alternative method mentioned in \cite{FCCMulticarrier}, which calculates the total bandwidth by summing the individual occupied bandwidths of each carrier frequency. Since we did not apply any modulation to the carriers, the sum of respective bandwidths will be extremely small, which can comply with the FCC regulation. Other modulated waveforms (\eg OFDM) cannot follow this alternative method and may potentially violate the regulation.

\section{Kernel-layer Combinations for Different Localization Algorithms}\label{sec: ToF and AoA}

Kernel-Layer near-field localization framework supports various localization algorithms because of the flexibility of measuring the similarity between receiving signal and theoretical signal and combining information across channels. For example, traditional ToF and AoA estimation algorithms can be implemented under the near-field condition with different kernels and layers.

\nosection{Kernel and Layers for ToF Estimation.}
ToF estimation can be done by choosing the following kernel and layer, where $\phi_{l}$ and $\theta_l$ are the empirical and theoretical phases at frequency $f_l$ respectively, and $d$ is the distance between tag and reader.

\begin{equation}\label{eqn: ToF layer}
  \begin{aligned}
        &\text{Kernel:}\  e^{-\boldsymbol{j} (\phi_{l} - \theta_l)}
         = e^{-\boldsymbol{j} (\phi_{l} - 2\pi f_l d/c)} = e^{-\boldsymbol{j}\phi_{l}} e^{2\pi f_l \tau}\\
        &\text{Layer:}\   \sum_{l=0}^{n} S(\tau) = \sum_{l=0}^{n} e^{-\boldsymbol{j}\phi_{l}} e^{2\pi f_l \tau}
  \end{aligned} 
\end{equation}

When using the above kernel and layer functions, $S(\tau)$ is the inverse Fourier transformation of the empirically measured phase value $\phi_1, \phi_2, ..., \phi_n$. Therefore, $S(\tau)$ is the time-of-flight expression of the empirically measured phases.

\nosection{Kernel and Layers for AoA Estimation.}\label{ssec:our hologram}
Similar to the ToF estimation, we can also design kernel and layer functions to extract angle-of-arrive (AoA) estimation. For the AoA estimation, we can use the following kernel and layer functions, where $\phi_{k}$ and $\theta_k$ are the empirical and theoretical phases at antenna k, respectively. $\Delta d$ is the distance between two neighboring antennas.

\begin{equation}\label{eqn: AoA layer}
  \begin{aligned}
        &\text{Kernel:}\  e^{-\boldsymbol{j} (\phi_k - \theta_k)}
     = e^{-\boldsymbol{j} (\phi_k - 2\pi f k \Delta d sin(\psi)/c)}\\
        &\text{Layer:}\  \sum_{k=1}^{m} S(\psi) = \sum_{k=0}^{m} e^{-\boldsymbol{j}\phi_k} e^{2\pi f k \Delta d sin(\psi)/c}
  \end{aligned} 
\end{equation}
$S(\psi)$ measures the similarity of the theoretical signal coming from angle $\psi$ and the empirically measured phase value $\phi_1, \phi_2, ..., \phi_m$ received by m antennas. Therefore, correct AoA $\psi$ is identified when $S(\psi)$ is maximized.

The summation layer, which sums up all the channels first by row and then by column, combines all the information for the final result. In this case, it combines near-field ToF and AoA estimations. We can develop more complex algorithms with the kernel-layer framework, such as the multipath-suppression algorithm in our paper.

\section{Direct Path Enhancement}\label{sec: time domain beamforming}

We enhance the direct path and suppress the influence from multipath with a frequency domain algorithm \cite{ma2017minding}. Assume there are $N$ paths with distances of $d_0,d_1,d_2,\dots,d_{N}$, and $d_0$ is the direct path. The channel $h_l$ of $l$th carrier can be expressed as:
\begin{equation}
    h_l = a_0 e^{-j\frac{2\pi}{c}f_l d_0} + \sum_{i=1}^{N} a_i e^{-j\frac{2\pi}{c}f_l d_i} \nonumber
\end{equation}
$a_i$ is the propagation attenuation of the $i$th path. To simplify the derivation without loss of generality, we assume $a_0 = a_i = 1,\ (i = 1,2,3,\dots)$, and what we measure is the phase of channel response:
$$ \phi_l = \angle h_l = \angle\{ e^{-j\frac{2\pi}{c}f_l d_0} + \sum_{i=1}^{N} e^{-j\frac{2\pi}{c}f_l d_i}\}$$

If we have a rough estimation of $d_0$, called $\tilde{d_0}$, we can use this algorithm to enhance the part of $a_0 e^{-j\frac{2\pi}{c}f_l d_0}$ (direct path) and suppress the part of $\sum_{i=1}^{N} a_i e^{-j\frac{2\pi}{c}f_l d_i}$ (multipaths) for a better location estimation.
In more detail, we use the prior knowledge of ROI to help determine the rough estimation of direct path $\tilde{d_0}$ with \algref{alg: direct path identification}. Then we enhance the direct path profile and suppress profiles of other paths by \eqnref{eqn: direct path enhancement} because the enhanced phase $\tilde{\phi_{l}}$ can be written as:

\begin{equation}
    \begin{aligned}
    \tilde{\phi_{l}}
    &= \angle\sum_{i=1}^{n} e^{j\phi_i}e^{j\frac{2\pi}{c}(f_i - f_l)\tilde{d}_{0}} \\
    &=\angle \{ e^{-j \frac{2 \pi}{c} f_{l} d_{0}} \sum_{i=1}^{N} e^{j \frac{2 \pi}{c}(f_i - f_l)\left(\tilde{d}_{0}-d_{0}\right)}\\
    &+\sum_{i=1}^{N} [ e^{-j \frac{2 \pi}{c} f_{l} d_{i}} \sum_{i=1}^{N} e^{j \frac{2 \pi}{c}(f_i - f_l)\left(\tilde{d}_{0}-d_{i}\right)}]\}\\ \nonumber
    \end{aligned}
\end{equation}

$\tilde{d}_{0} \approx d_0$ so $(\tilde{d}_{0} - d_0)\Delta f/c \ll 1$, and it leads to:
\begin{equation}
\sum_{i=1}^{N} e^{j \frac{2 \pi}{c}(i-l) \Delta f\left(\tilde{d}_{0}-d_{0}\right)} \approx \sum_{i=1}^{N}1 = N  \nonumber
\end{equation}
For multipath whose $d_i$ is different from $\tilde{d}_{0}$, $\tilde{d}_{0} - d_i$ is large so
\begin{equation}
\left|\frac{\sum_{i=1}^{N} e^{j \frac{2 \pi}{c}(f_i - f_l)\left(\tilde{d}_{0}-d_{i}\right)}}{N}\right| \approx\left|\operatorname{sinc}\left[B\left(\tilde{d}_{0}-d_{i}\right) / c\right]\right| \ll 1 \nonumber
\end{equation}

The part of the direct path is much larger than the part of other paths, so the direct path is reinforced. $\tilde{d}_{0}$ helps to get rid of the leakage interference from multipath, and the following summation layer can make a better estimation of $d_0$ as the final output. 
Besides using the prior knowledge, other methods (\eg fingerprinting-based algorithm, Bayesian-based algorithm) can also be used to determine the rough estimation $\tilde{d_0}$, which is beyond the scope of this paper.